# NMR contributions to the study of water transfer in proton exchange membranes for fuel cells

Jean-Christophe Perrin*, Assma El Kaddouri, Laouès Guendouz, Christine Mrad, Kévin Mozet, Jérôme Dillet, Sébastien Leclerc, Olivier Lottin

*Université de Lorraine, CNRS, LEMTA, F-54000 Nancy, France*

** corresponding author:* jean-christophe.perrin@univ-lorraine.fr (J.-C. Perrin)

**Abstract**

As programs to support efficient and sustainable energy sources are expanding, research into the potential applications of the hydrogen vector is accelerating. Proton exchange membrane fuel cells are electrochemical converters that transform the chemical energy of hydrogen into electrical energy. These devices are used today for low- and medium-power stationary applications and for mobility, in trains, cars, bicycles, etc. Proton exchange membrane fuel cells use a polymer membrane as the electrolyte. The role of the membrane is multiple: it must separate gases, be an electronic insulator and a very good ionic conductor. In addition, it must resist free-radical chemical attack and have good mechanical strength. Nafion-type perfluorinated membranes have all these properties: the fluorinated backbone is naturally hydrophobic, but the hydrophilic ionic groups give the material excellent water sorption properties. The water adsorbed in the structure is extremely mobile, acting as a transport medium for the protons generated at the anode. Although it has been studied for a long time and has been the subject of a large number of papers perfluorinated membranes are still the reference membranes today. This article reviews some contributions of Nuclear Magnetic Resonance methods in liquid state to the study of water properties in the structure of Nafion-type perfluorinated membranes.

# 1. Introduction

## 1.1. General information on PEMFC

A proton exchange membrane fuel cell (PEMFC) is an electrochemical converter that combines hydrogen and oxygen to produce electrical energy, water, and thermal energy. Among different types of fuel cells, PEMFC technology currently dominates the market due to its suitability for portable applications ($< 1\ kW$) and transportation sector ($\sim 100\ kW$ for automotive vehicles). To achieve these power levels, one or several stacks of individual cells are used, with each cell capable of producing up to $1\ W/cm^2$ of active surface area. The operational principle of a single cell is relatively simple, but its engineering is rather complex. The various components must possess the required properties to fulfill the necessary requirements for proper functioning, including uniform supply of reactants, separation of anode (where the hydrogen oxidation half-reaction occurs) and cathode (where the oxygen reduction half-reaction occurs) compartments, charge transport across the membrane-electrode assembly, and evacuation of produced water without impeding reactant transport. The different components constituting a PEMFC cell are as follows (

*Figure 1*):

- **The membrane-electrode assembly** (MEA), which consists of electrodes where electrochemical reactions take place, and is made out of carbon black, platinum nanoparticles, and ionomer. The electrodes must support the catalyst, have good electronic and ionic conductivity, enable the diffusion of reactive gases to reaction sites, and facilitate water evacuation.

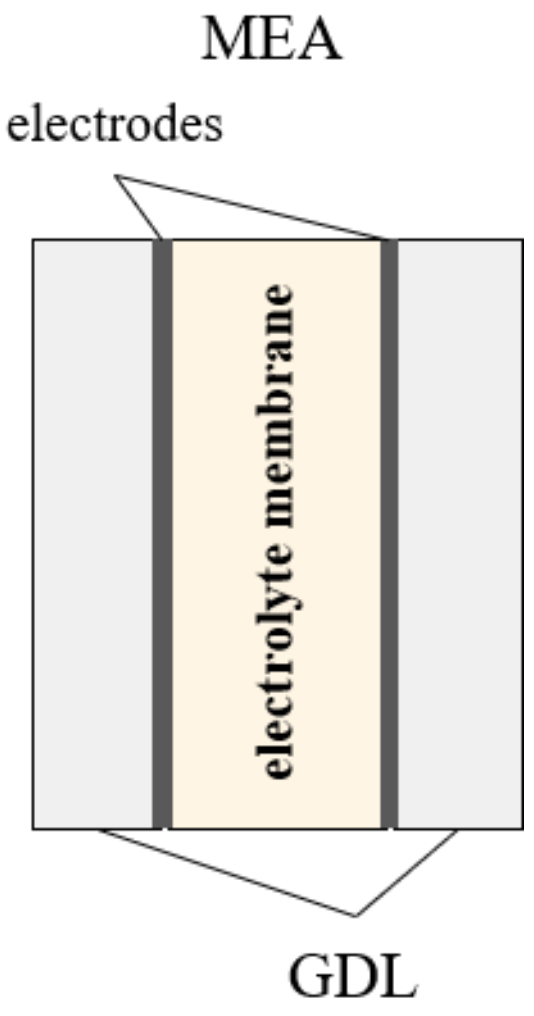

Figure 1.

Main components of a PEMFC cell.

- **The ionomer electrolyte membrane**, at the center of the cell. Currently, the membrane has a thickness of less than 15 $\mu m$ and ensures efficient transport of protons from the anode to the cathode as well as water diffusion from the cathode to the anode, while providing electronic insulation between the two compartments.
- **The gas diffusion layer** (GDL), which enables uniform distribution of reactants to the electrodes and facilitates water management within the cell through hydrophobic treatment. Its thickness ranges from 200 to 400 $\mu m$, with pore sizes of approximately 10 $\mu m$. A microporous layer, typically more hydrophobic with smaller pores, can be placed between the electrode and the GDL to enhance electrical contact and limit electrode flooding by liquid water.
- **The flow field plates** that distribute reactants and remove excess water and gases. They also ensure compression of the different layers and allow for the collection of electrical current.

PFSA membranes consist of a fully fluorinated polymer chain to which are grafted fluorinated pendant chains terminated by a sulfonic acid group. Nafion is the best-known PFSA polymer. It is composed of a hydrophobic fluorocarbon backbone derived from polytetrafluoroethylene -PTFE-. The pendant chains are terminated by hydrophilic sulfonate groups ($-SO_3^- -$), neutralized by a counterion ($H^+$ in acid medium). Currently sold by the Chemours Company, it was introduced in the 1960s by DuPont, then used in numerous industrial processes, such as production of sodium hydroxide by brine electrolysis, or more recently, in sensors (humidity sensors, biosensors). Nafion has been studied in laboratories and industry for about 60 years (*1*).

### 1.2. Water in PEM membranes

Water plays a fundamental role in the operation of a PEMFC: it is the product of the electrochemical reaction, the material that fills the membrane's porous microstructure, and the transport medium for protons transferred from the anode to the cathode. While the polymeric backbone provides the membrane's mechanical properties, the sulfonic acid function confers its water sorption and swelling properties. Conventionally, the amount of water present in the polymer structure is quantified by the parameter $\lambda$, representing the average number of water molecules adsorbed on each sulfonic acid unit:

$$\lambda = \frac{[H_2O]}{[SO_3H]} \qquad \text{equation 1}$$

In Nafion membranes, $\lambda$ varies between $0$ and around $10$ to $15$ when the air relative humidity rises from $RH = 0\%$ to $RH = 100\%$, while the relative change in membrane volume (swelling) is significant, typically $\sim 15\%$, between the dry and saturated states. The "performance" of an ion-conducting membrane is quantified by its ionic conductivity, measured in $S.cm^{-1}$. The more efficient the ionic transport, the better the membrane material performs as an electrolyte. At a given temperature, the proton conductivity of ionomer membranes varies tremendously as a function of water content (*2, 3*). As water is essential for proton transport, the study of water diffusion through the membrane is crucial for PEMFC applications. The mechanisms of proton transport in Nafion have been studied and debated in the literature (*4*). It is accepted that there are two main modes of conduction: the mass diffusion of hydronium ions $H_3O^+$, also known as the vehicular mechanism, and the transfer of protons via hydrogen bonds from one vehicle to another. These processes are strongly coupled to the reorganization of the environment: both the migrating species and the solvent. A third mechanism, surface diffusion, has also been proposed at low hydration. The correlation between the mechanisms depends on the state of water confinement and therefore on the degree of hydration of the membrane. In these materials, it is the simultaneous existence of ionic functionalities and a hydrophobic backbone that is at the origin of the nano-phase separation between hydrophilic and hydrophobic domains, which generates a network of multi-connected "pores" accessible to water. In the case of Nafion, numerous and sometimes competing structural models have emerged since the 1980s. All studies show that the structure is complex and organized at different spatial scales. In the model from Rubatat et al. (*5, 6*), the basic brick is an aggregate of cylinder- or ribbon-shaped polymer chains. These elongated objects are arranged in bundles to form organized zones with a characteristic size of about 500 Å (*Figure 2.*). Finally, the bundles are organized isotropically on a larger scale.

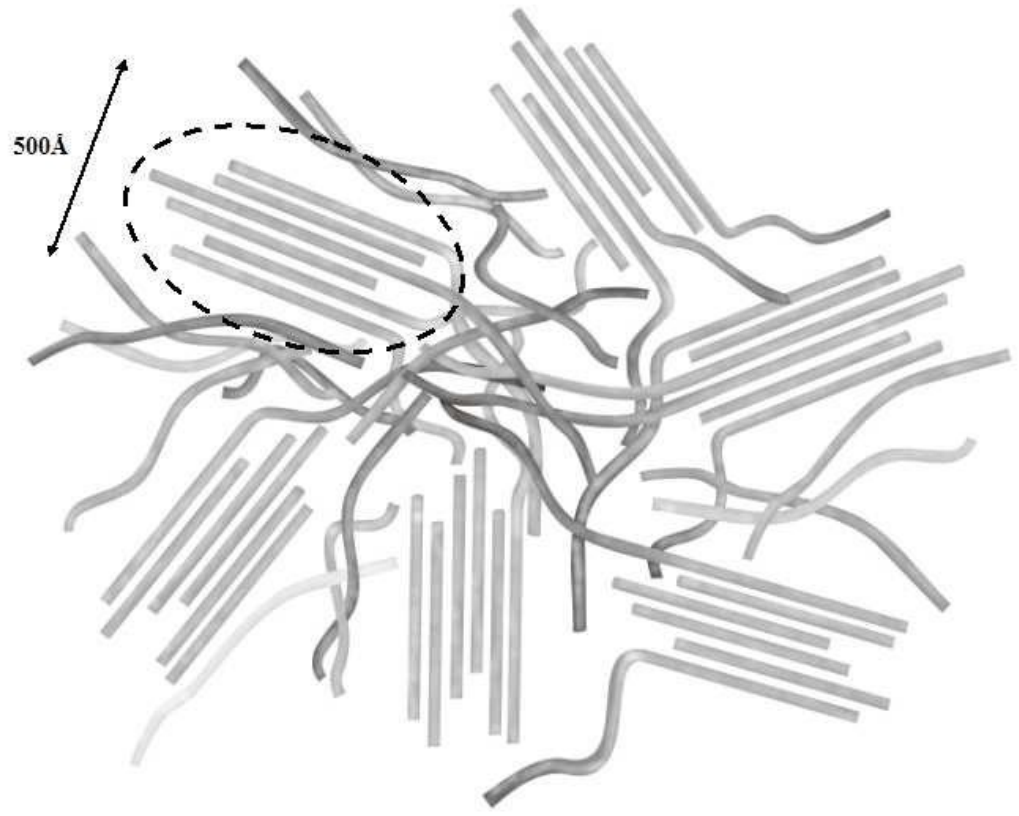

Figure 2.

Diagram of several "cylinder bundles" in Nafion. This drawing shows very schematically organized zones with a characteristic size of around 500Å (circled in dotted line) and completely disordered zones. (the parallelism between the cylinders is exaggerated). Figure from ref (*6*).

Hydrated membranes are mixed polymer/water matrix systems containing particularly complicated confinement matrices, since the interaction of water and charges, swelling and the multi-scale nature of the structure have to be considered. In such structures, the water diffusion coefficient depends on the time scale probed by the experimental technique used. The molecular scale, studied by quasi-elastic neutron scattering, reveals rapid diffusion over a characteristic time of the order of picoseconds (*7*). In contrast, self-diffusion probed by pulsed-field gradient NMR (PGSE-NMR) methods reveals a diffusion coefficient averaged over a time of the order of milliseconds (*8*). Between the two, water diffusion undergoes slowdowns at the so-called "intermediate" scale, that can be probed by field cycling NMR relaxometry (*9-11*).

## 2. NMR study of water properties in PFSA membranes

From the earliest days, NMR was used to study diffusion in liquids (*12*). From 1965 onwards, with the application of pulsed field gradient methods by Stejskal and Tanner (Stejskal and Tanner 1965), the magnetic field gradient intensity could be chosen strong enough to allow measurement of diffusion in porous media. The methods involved are sensitive to molecular motion and therefore reveal information about porous structures that impede molecular movement (*13, 14*). NMR methods have also been used for a long time to study fuel cell materials, particularly membranes, and can even be said to have played a part in the development of PEMFC. NMR has been extensively exploited because the methods are well suited to the study of molecular motion in fuel cell membranes. Water is abundant and the phenomena to be studied are numerous and sufficiently complex to mobilize a large number of researchers (*15*). In PFSA membranes, the situation is ideal, as proton NMR only provides information on adsorbed water, while fluorine NMR can probe the chemical structure of the perfluorinated skeleton. In hydrocarbon materials, the situation is more complex, but filtering of the matrix signal is relatively straightforward, so that the water signal is accessible. The review article by Yan et al. (*16*) details in this sense the contributions of the different techniques to research on core materials. The article by Zhang et al. (*17*) reviews studies using MRI as a "diagnostic tool" for fuel cells.

### 2.1. proton NMR spectra in PFSA membranes.

The main feature of proton NMR spectra of Nafion-type PFSA membranes is an intense resonance line of liquid nature, broadened by the resolution of the spectrometer (*Figure 3*). This line is generally detected at a chemical shift between 6.5 and 12 $ppm$ relative to the TMS reference and is attributed to the water present in the ionic domains. First, no significant broadening occurs as the water content of the membrane decreases, although the transverse relaxation time $T_2$ drops significantly.

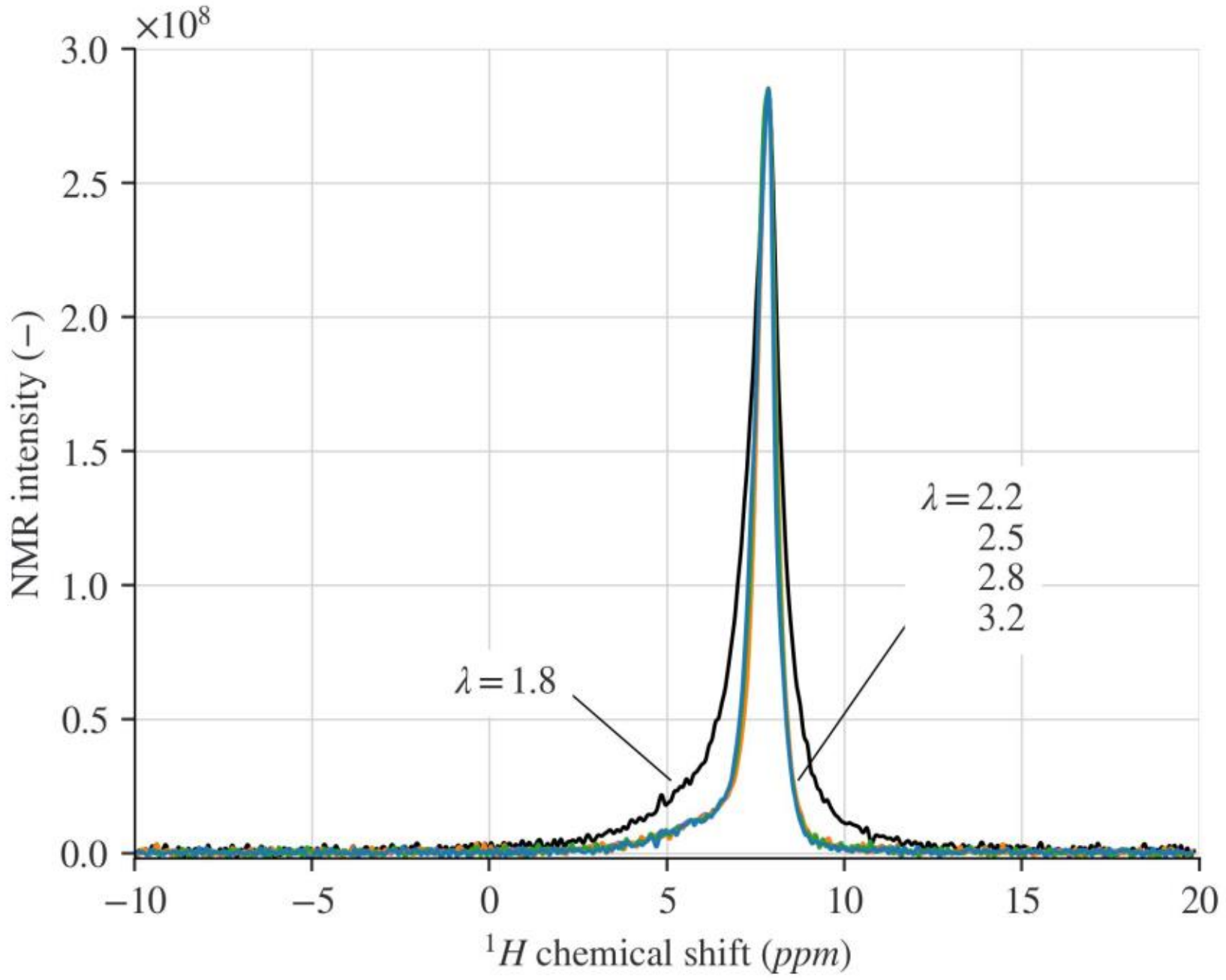

Figure 3. [1]H NMR spectra of low-water content Nafion 112, measured at $200\,MHz$, $25°C$, and normalized to the maximum intensity. Adapted from reference (*18*).

A broadening appears only at very low water content ($\lambda < 2$), indicative of a significant reduction in molecule mobility. In all cases, the line shape is highly asymmetrical, indicating the existence of one or more other components, albeit unresolved. These additional components may correspond to protons that cannot be exchanged with those of the main water population on the characteristic time scale of the measurement: $H^+$ counterions or impurities introduced by synthesis. As the amount of water in the membrane increases, the main resonance line shifts towards the low chemical shifts ($\delta$) (*19, 20*): the network of hydrogen bonds becomes denser, the water interacts less and less with the polymer chains, and becomes increasingly mobile. This dependence of $\delta$ on $\lambda$ inspired Bunce et al. (*19*) to develop a method for determining the absolute amount of water in a Nafion membrane through the direct measurement of $\delta$. The problem of choosing a reference state arose immediately: it is known that it is practically impossible to remove all the water from a PFSA membrane other than by heating it to high temperatures (*21*).

### 2.2. NMR relaxation and water mobility

The interactions between the proton magnetic moments in a sample can often come from a variety of sources. The measurement of the proton NMR relaxation parameters ($T_1$, $T_2$) provides information on the origin of the modulations, i.e., on the molecular dynamics of the spin carriers, by revealing the nature of the modulations of the interactions between the spins in the system and, consequently, the nature of the relaxation mechanisms. In the case of dipolar interaction between spins with random rotational motion, the BPP model (*22*) shows that the correlation function is an exponential of the form $exp(-t/\tau_c)$, where $\tau_c$ is a correlation time characteristic of the motion. Two modulation ranges can be distinguished as a function of this time: a fast regime -also called "extreme narrowing"- when the correlation time is much less than the inverse of the Larmor frequency, ($\omega_0\tau_c << 1$) and a slow regime if the correlation time is much higher than this value, ($\omega_0\tau_c >> 1$). In the fast modulation regime, the relaxation rates $R_1$ and $R_2$ are equal and proportional to $\tau_c$. In Nafion, the time evolution of nuclear magnetization recorded during the so-called CPMG measurement (used to measure $T_2$) and inversion-recovery measurement (to measure $T_1$) is systematically mono-exponential, characteristic of rapidly exchanging water populations, on the characteristic time scale of NMR acquisition. The evolution of the relaxation rates measured at room temperature as a function of the water content in a Nafion 112 membrane is plotted in Figure 4. The values of $R_2(\,^1H) = 1/T_2(\,^1H)$ were measured at $20\,MHz$ with the CPMG sequence. The transverse water relaxation times $T_2$ are long ($> 50\,ms$ for $\lambda > 3$), with a variation of more than two decades over the entire hydration range. For large values of $\lambda$, $R_2$ approaches $R_1$ to reach the conditions of extreme narrowing of liquid water at $1/\lambda = 0$ for which $R_1(200\,MHz) \approx R_2(20MHz)$. We also observe that bulk water appears as the infinite dilution limit of $R_1(\lambda)$ for $1/\lambda \rightarrow 0$. The evolution of $R_1$ at "high" field indicates that $R_1(\lambda \rightarrow 0)$ reaches the maximum value of the relaxation rate at $\omega_0/2\pi = 200\,MHz$. This property is representative of molecular dynamics with a characteristic time $\tau_c$ such that $\omega_0\tau_c = 1$, i.e., $\tau_c \approx 1ns$ a value well below that of bulk water ($\approx 1ps$).

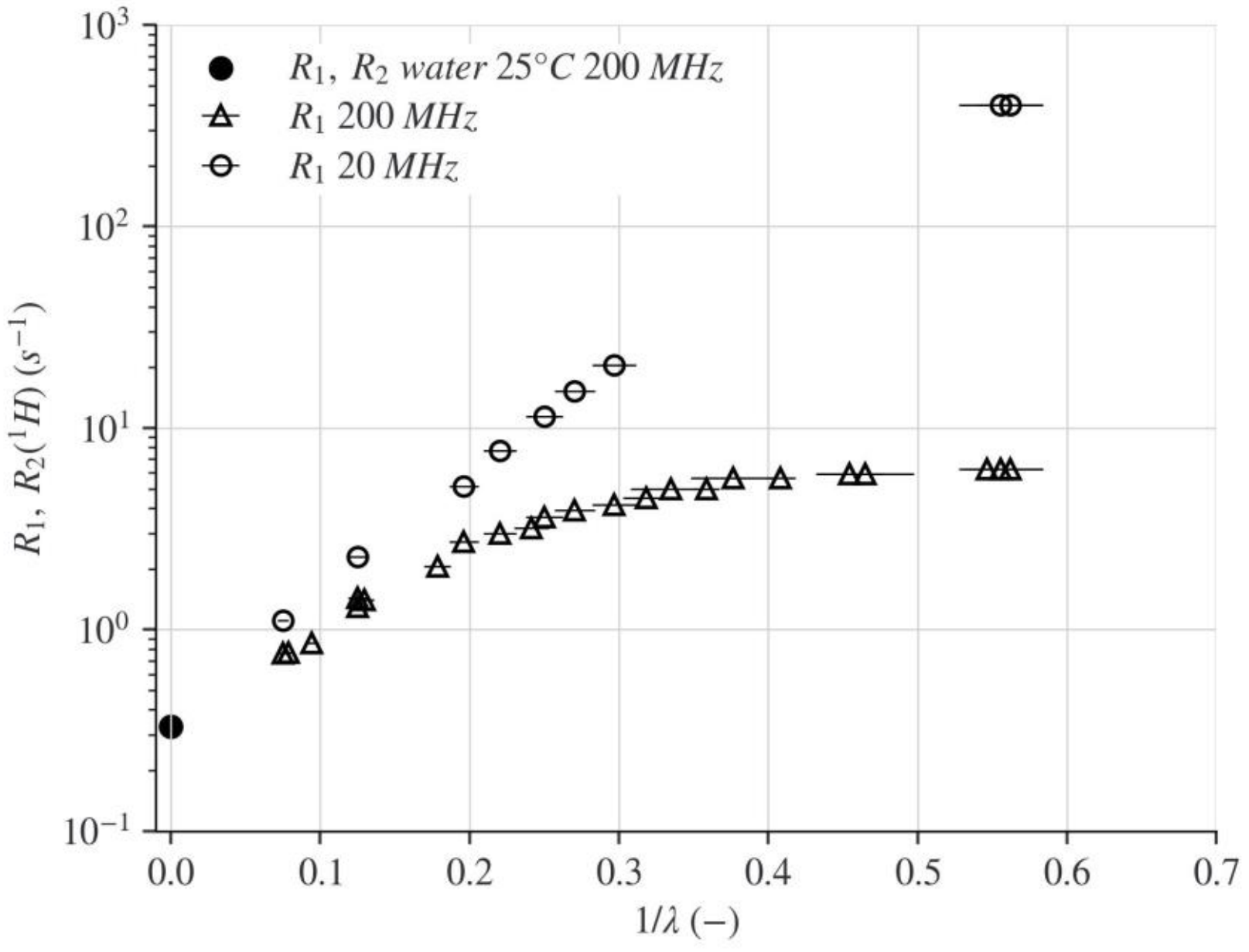

(a)

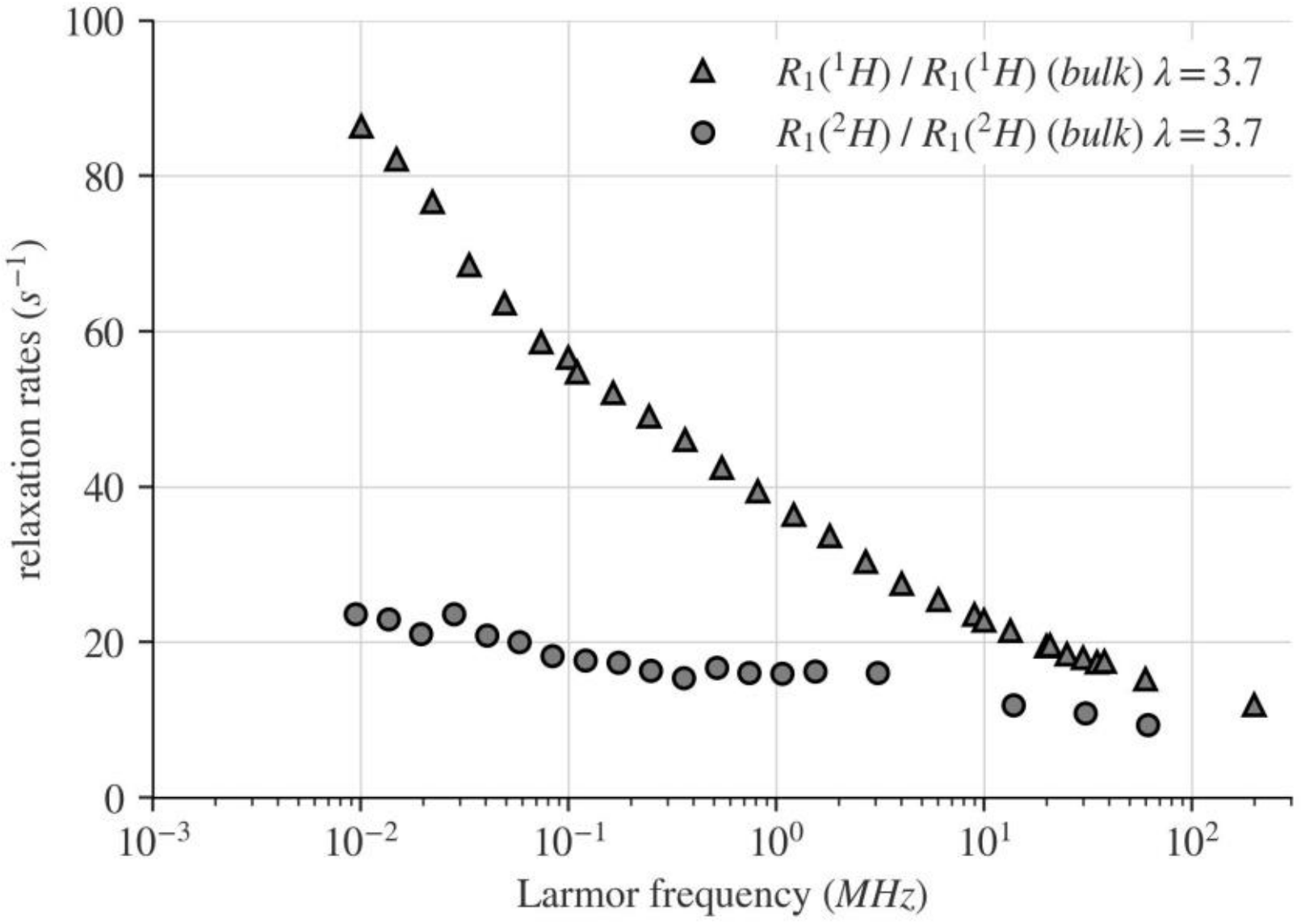

(b)

Figure 4

(a) Proton relaxation rate in a Nafion 112 membrane at room temperature as a function of $1/\lambda$. Adapted from reference (*18*). (b) Normalized proton and deuteron relaxation profiles in Nafion 112 at $\lambda = 3.7$. Adapted from reference (*18*).

To go one step further, temperature and Larmor frequency can be used as experimental parameters to study dynamics on "short" or "intermediate" time scales, respectively. In 1999, MacMillan et al. produced a comprehensive study of the restriction of molecular motion by the presence of "pores", by measuring the temperature evolution of proton, deuteron, and fluorine relaxation times in hydrated Nafion membranes with different water contents (*23*, *24*). The authors have identified a transition temperature, above which the water in the clusters behaves like bulk water. Below this temperature, the slopes of the $T_1(T)$ and $T_2(T)$ curves are much steeper, indicative of a situation in which molecular movement is restricted. The transition temperature depends on the water content of the membrane, and ranges from $-10°C$ to $13°C$ for high and low water contents respectively. The authors note that this transition does not correspond to a liquid ↔ solid phase transition, which was later confirmed by further relaxation and diffusion studies (*25*, *26*). Furthermore, no structural transition is possible since for the same temperature ranges, no discontinuities are observed in the fluorine relaxation curves. An important finding of the MacMillan et al. study is that, in the high-temperature limit ($T > 270°K$), $T_{1\rho} = T_2 \neq T_1$, which is unusual behavior and not understandable within the framework of the BPP model (*22*). This model predicts that, in the so-called "extreme narrowing" regime, the three relaxation times $T_1, T_2$ et $T_{1\rho}$ converge to a single value. Since $T_1$ represents the longitudinal relaxation time at high frequency, $T_{1\rho}$ the longitudinal relaxation time at low frequency, and since $T_{1\rho} = T_2$ in this temperature range, the authors attribute the relatively short values of $T_1(\ ^1H)$ to the presence of an additional relaxation mechanism at low frequency. In Nafion, the movement of confined water is slowed down by interactions with the material's surfaces. Processes appear that are much slower than molecular movements in the bulk: several elementary steps are then required for a molecule to lose the correlation of its initial orientation or position. The additional relaxation mechanism suggested by MacMillan et al. has been identified and studied using the NMR relaxometry method (field-cycling NMR). In this method, the Larmor frequency at which the nuclear spins relax is used as a parameter (*27*) to extract information on the slow movements of water molecules: typically from $10\ \mu s\ to\ 1\ ns$ in the membranes studied here, corresponding to Larmor frequencies between $10\ kHz$ and $400\ MHz$. The shape and amplitude of the dispersions are indeed characteristic of modulations by spin carrier dynamics of the interactions responsible for relaxation. To be exploitable, however, the data must be interpreted within the framework of theoretical models, which are only valid if the mechanisms behind nuclear relaxation are clearly identified. Proton relaxation is governed by the dipolar interaction. In water, a proton is in dipolar interaction with the second proton of the same molecule and with other neighboring protons belonging to different molecules. Intramolecular dipolar interactions (between protons of the same molecule) are modulated by molecular reorientations, while intermolecular dipolar interactions are modulated by variations in the radius-vector joining the protons, i.e. by translational diffusion. Measuring the proton relaxation time alone is therefore not enough to directly assess the respective proportion of relaxation due to the intra- and inter-molecular components of the dipolar interaction. In the case of deuteron, on the other hand, relaxation is largely due to rotational modulation of the intramolecular coupling between the deuteron's quadrupole moment and the electric field gradient carried by the $O-D$ bond. This interaction is strong, and the dipole-dipole interaction is negligible. The measurement of $T_1(\,^2H)$ can therefore be directly interpreted in terms of rotational motion. The two contributions, rotational and translational, can be distinguished and separated by measuring the two relaxation times $T_1(\,^1H)$ and $T_1(\,^2H)$.

The main conclusions of NMR relaxometry studies are as follows (*18*):

- The evolution of proton relaxation rates as a function of Larmor frequency measured over the entire hydration range shows little variation, a sign that the interaction between water and the polymeric surface is not very intense.
- The proton and deuteron relaxation dispersions, normalized to the relaxation rates of water and heavy water (compared in Figure 4-(b) for low hydration), are very distinct, meaning that molecular reorientation modulating the intramolecular dipolar interaction is not the main contribution to relaxation. The difference between the two profiles proves that there is at least one other relaxation mechanism, corroborating the findings of MacMillan's study.
- Proton relaxation contains an intramolecular component that can be evaluated by relaxation measurements on heavy water (at identical hydration) and subtracted. The resulting intermolecular contribution contains a component due to the interaction between water protons and fluorine nuclei in the matrix, and another due to the modulation by translational motion of water molecules of the $H-H$ dipolar interaction between protons belonging to different molecules. It has been shown that the $H-F$ dipolar interaction is in the minority at low frequencies and for water contents above $\lambda \approx 6$. In this hydration range, all dispersions follow a logarithmic distribution. They were then interpreted within the framework of a dipolar relaxation model by translational diffusion in a locally two-dimensional medium (*28*). For higher water contents, the relaxation of water protons is almost exclusively due to intramolecular components.

In conclusion, we can state that the non-wetting nature of the perfluorinated Nafion membrane induces relatively low-dispersion relaxation profiles. The transition from $2D$ to $3D$ diffusion during hydration is a realistic hypothesis for interpreting the evolution of intermolecular dipolar interaction relaxation profiles as a function of water content. This transition takes place around $\lambda = 5 - 6$. These results support a structural model of Nafion based on the existence of elongated aggregates, organized in locally lamellar domains at the nanometric scale. A small increase in the size of the inter-lamellar space, due to the introduction of additional water, induces a rapid disappearance of the $2D$ behavior and the establishment of the $3D$ regime, which is directly reflected by the significant variation in dispersion profiles observed over a restricted hydration range, from $\lambda = 3.7$ to $\lambda = 5.6$.

### 2.3. PFGNMR and diffusion at the micrometer scale

In a pulsed field gradient diffusion (PFG-NMR) measurement, motion detection is based on the analysis of the attenuation of the NMR signal, known as spin echo, resulting from the dephasing of nuclear spins under the combined effect of translational motion and the application of perfectly controlled field gradient pulses (*13*). In contrast to the relaxation methods described in the previous paragraph, the spatial scale associated with the measurement is directly linked to the duration of motion observation set by the user (diffusion delay $\Delta$). In the case of free diffusion, the value of the self-diffusion coefficient does not depend on the observation time, as the root-mean-square displacement is a linear function of time. As the self-diffusion coefficient of free pure water at $25°C$ is around $2.3.10^{-9}\ m^2.s^{-1}$ (*29*), the scale probed along the direction of application of the field gradient in this case is de $\sqrt{2D_s\Delta} \approx 5$ to $25\ \mu m$ for $\Delta = 5$ to $150\ ms$. In the case where the fluid (water for example) is confined, the displacement will depend on the diffusion delay, the size of the confinement domain and the bulk diffusion coefficient. In the Nafion membrane, the PFG-NMR method has been used since the 1990s to study the link between water sorption, molecular diffusion, and ionic conductivity (*30, 31*). More recently, other studies have used this method to elucidate proton transport mechanisms (*32*). The values of the water diffusion coefficient $D_s$ increase strongly with water content, in the range $2.10^{-8} < D_s < 6.5.10^{-6} cm^2/s$ (*8, 21, 33*). The value in the saturated membrane is lower than that measured for pure liquid water at the same temperature by a factor $\sim 4$ $(23.10^{-6} cm^2/s)$. This may be related to the tortuosity of the medium (Figure 5).

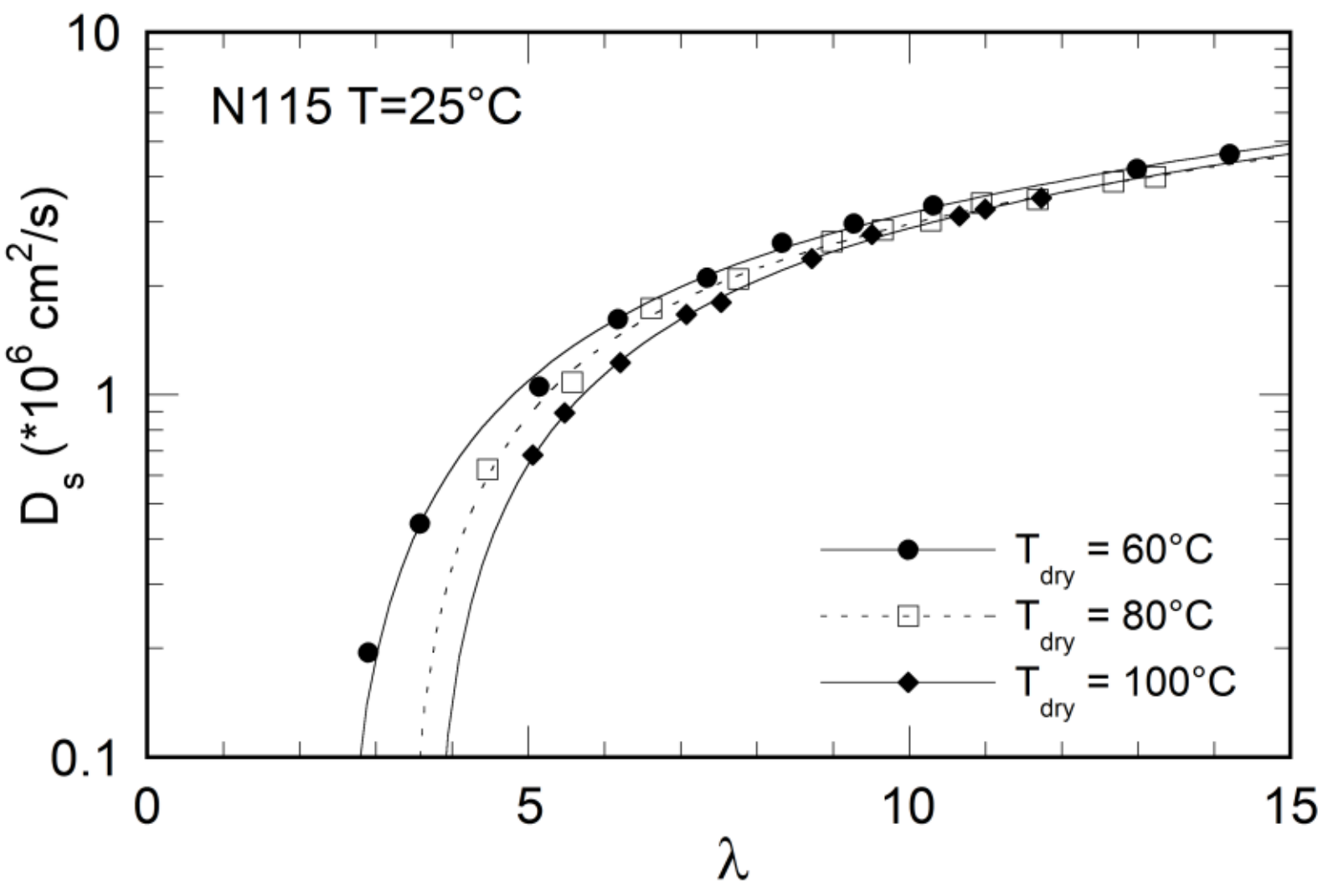

Figure 5. Influence of water content $\lambda$ and drying temperature on the evolution of the self-diffusion coefficient and proton conductivity in Nafion N115. From reference (*21*).

In general, diffusion is activated by temperature. Assuming that the diffusion coefficient follows an Arrhenius law, the activation energy extracted varies between 31.6 and 19.2 $kJ/mol$ as $\lambda$ increases from $\sim$3 to $\sim$13. By comparison, the activation energy extracted from the Arrhenius plot of the diffusion coefficient of liquid water over the same temperature range is 16.7 $kJ/mol$. The continuity between the values in Nafion and liquid water is good; diffusion in liquid water then appears as the infinite dilution limit of the diffusion of water in Nafion. Water in Nafion thus acquires the same diffusive behavior as free water when hydration is at its maximum. However, as hydration decreases, the energy cost becomes much higher.

### 2.4 Impact of chemical degradation

The degradation and ageing of PEMFC materials have been the target of numerous experimental and theoretical studies (*34*, *35*). Membrane aging and the alteration of its properties is a particularly important issue. During the operation of a PEMFC, the membrane undergoes significant modifications, of various origins (mechanical, chemical, electrochemical and/or hygrothermal), which can affect the performance of the cell, or even lead to its shutdown. The problem of membrane ageing is difficult to solve, as it is the consequence of simultaneous and coupled mechanisms. It must therefore be tackled using a multi-analytical experimental approach. The effects induced by the use of a Nafion XL membrane in a fuel cell stack at constant intensity over a long period have been studied by proton and fluorine NMR (*36*). An example of the results is shown in Figure 6. The XL membrane is a tri-layer composite membrane with a PFSA-impregnated PTFE microporous layer in the center. Nominal thicknesses are $\sim 12\ \mu m$ for the PTFE layer and $\sim 9\ \mu m$ for each PFSA outer layer. The samples studied came from an Axane Evopac system comprising two 55-cell stacks, operated under real-life conditions at $0.26\ A/cm^2$ for $12860\ h$, with around 250 start/stop sequences (*37*, *38*). During operation, the voltage of the cell containing the membrane studied here dropped from $0.718\ V$ to $0.706\ V$.

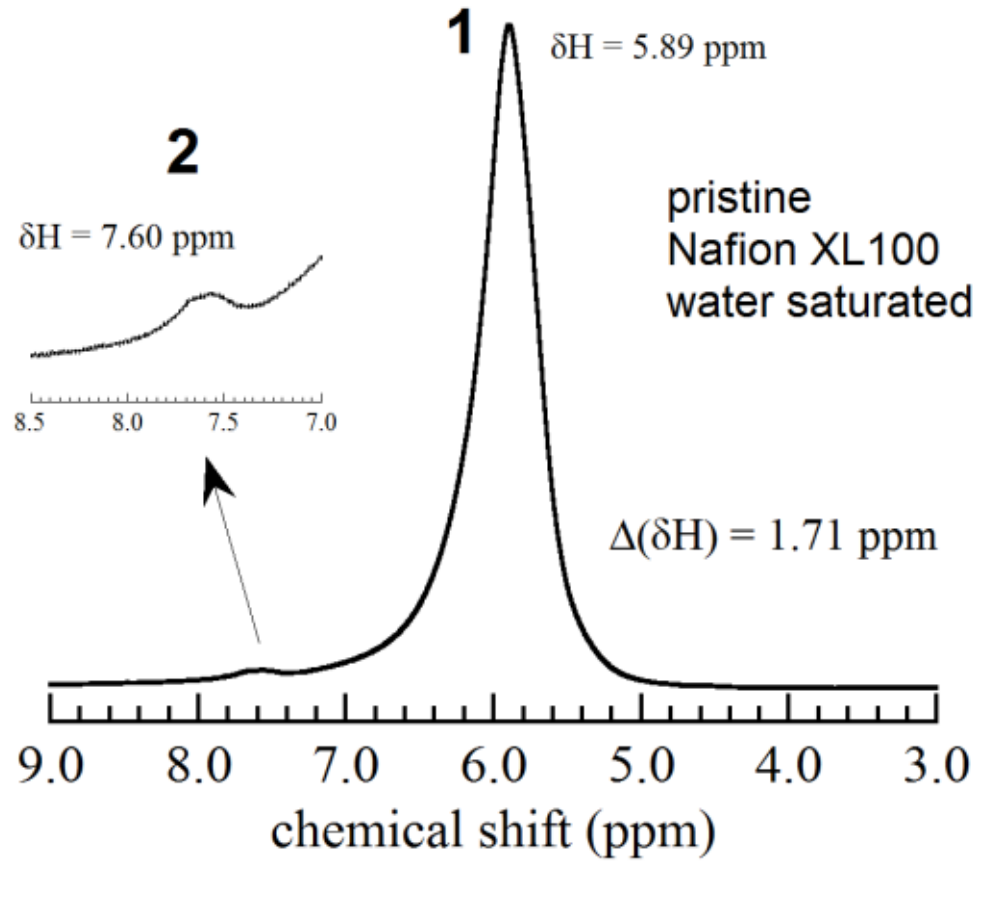

(a)

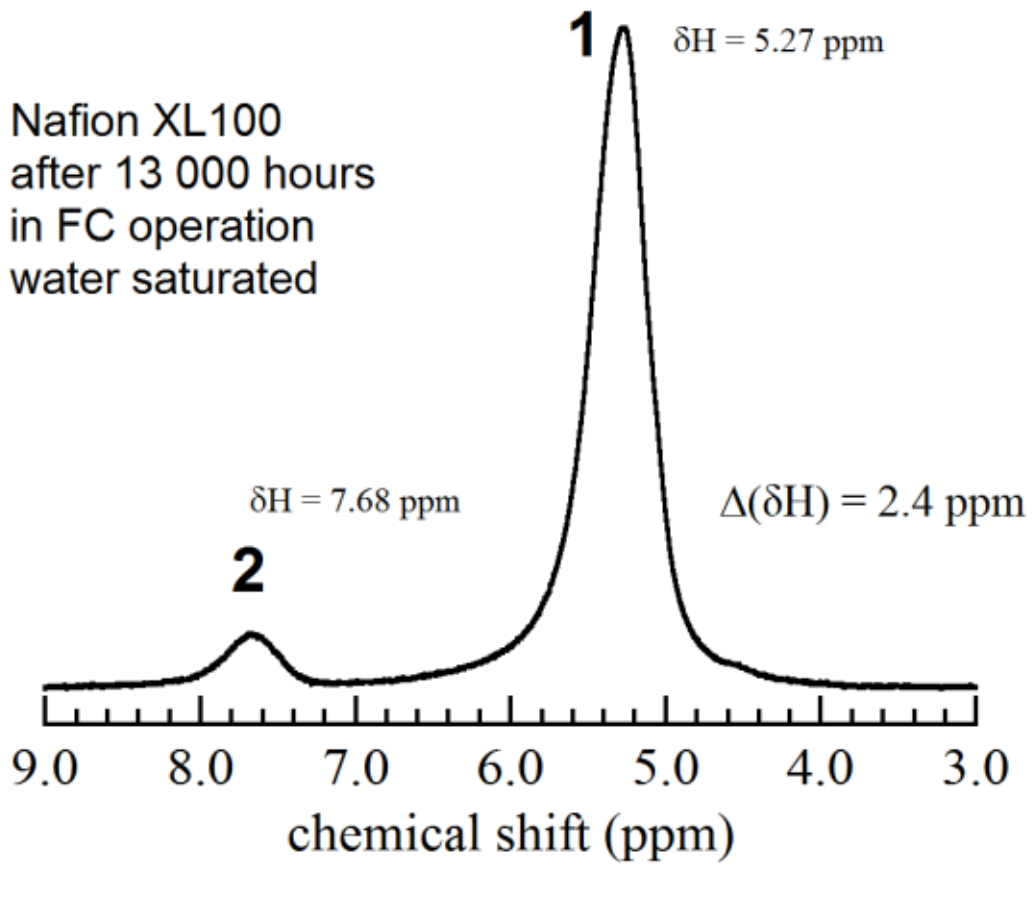

(b)

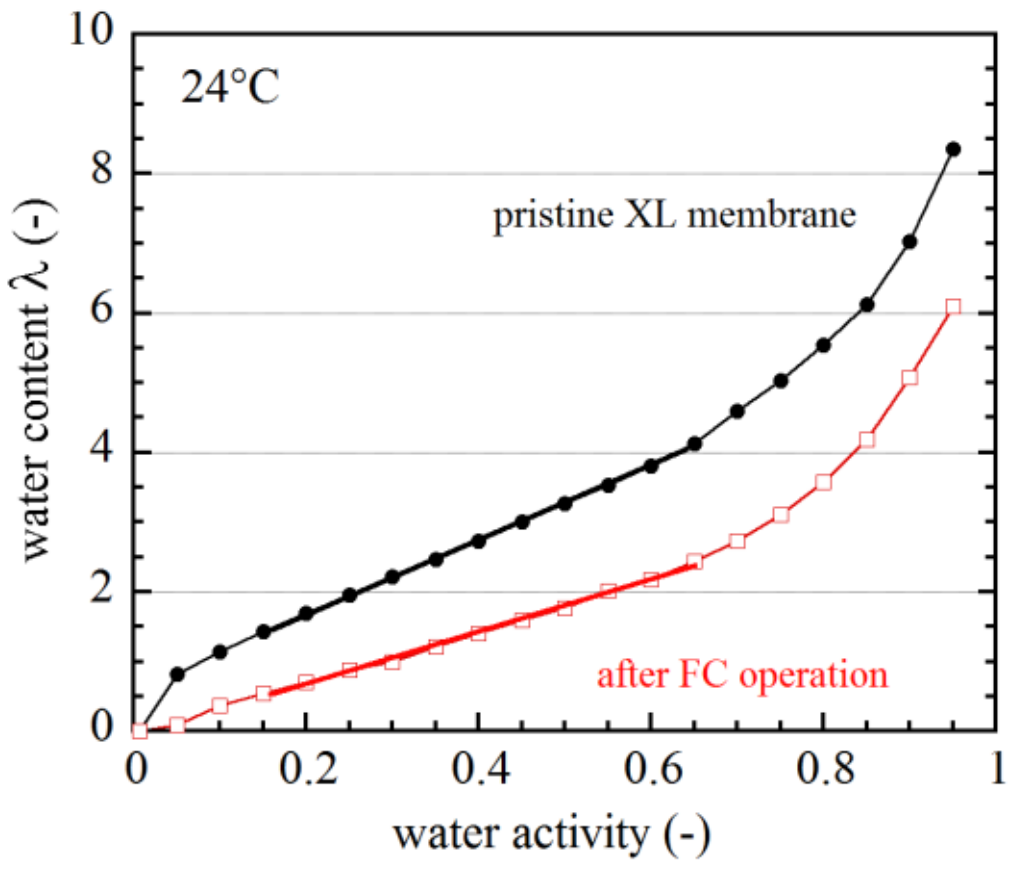

(c)

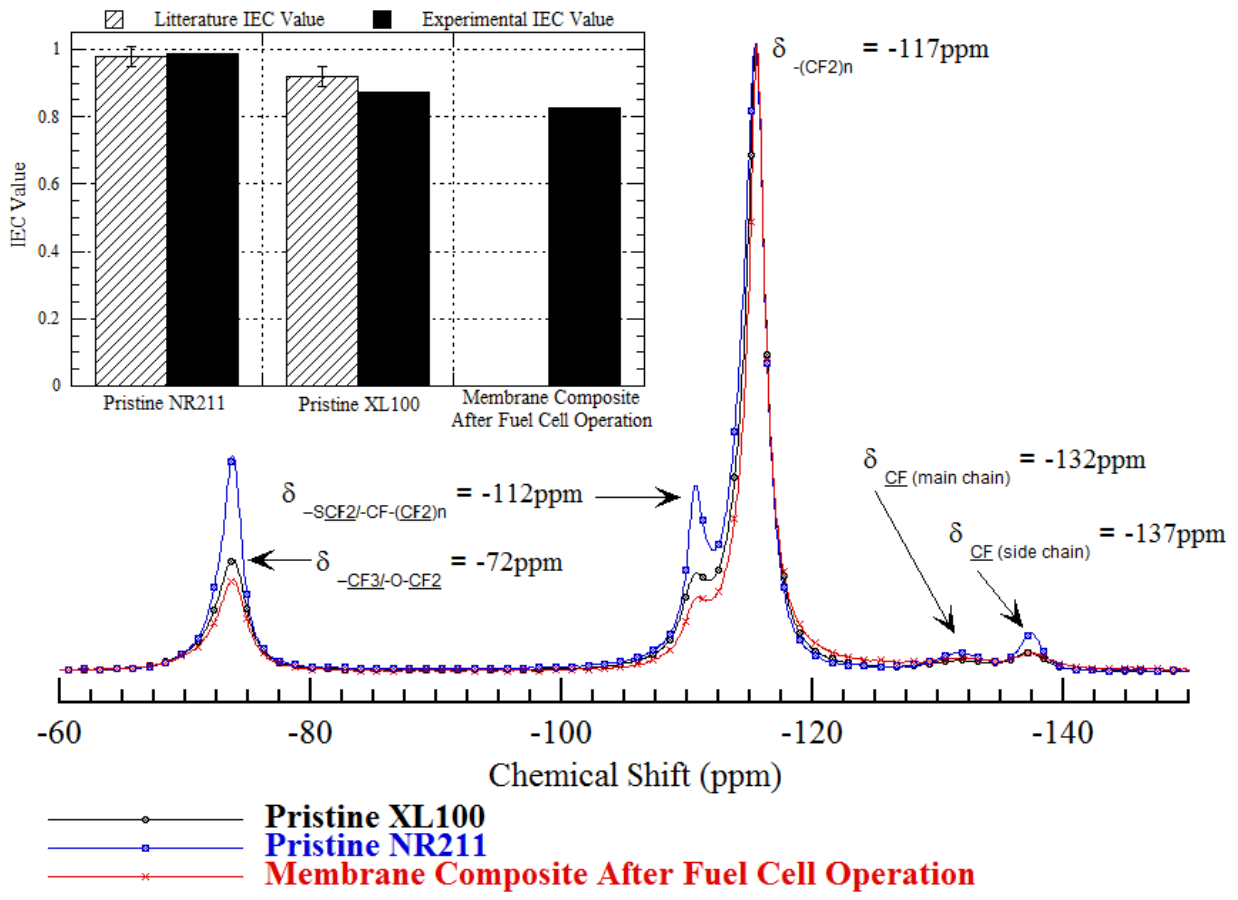

(d)

Figure 6. (a) and (b) $^{1}H$ NMR spectra of Nafion XL100 membranes before and after PEMFC operation. (c) Comparison of membrane sorption isotherms before and after operation. (d) $^{19}F$ NMR spectra of new XL100 and NR211 membranes and the tri-layer membrane after operation. Adapted from (*36*).

The $^{1}H - NMR$ spectrum of the new XL100 membrane shows a second resonance peak at around $7.6\ ppm$. Pulsed field gradient and relaxation experiments show a complex behavior, with (at least) two components (*36*). Initially attributed to the water present in the PTFE reinforcement, this second population could correspond instead to hydronium ion protons, an attribution made by Han et al. in Nafion NR211 (*39*) (the NR211 membrane is a monolayer, with a nominal thickness of 25 $\mu m$, based on chemically stabilized PFSA polymer). After fuel cell operation, the sorption curve shows a reduced water uptake capacity, while the resonance line attributed to water shifts towards the lowest chemical shifts, which is contrary to the behavior observed in new membranes: the water's chemical environment has therefore been modified. The self-diffusion coefficient, meanwhile, follows the usual logic, with a lower value in the saturated membrane after operation (containing less water than the new membrane). The $^{19}F$ spectra show that operation has altered the polymer, as the ion exchange capacity (denoted IEC in the figure) is lower after operation. This may be due to the partial loss of ionic sulfonic groups as a result of side-chain scissions caused by radical attack (*40, 41*) (Figure 7). Indeed, it is known in the literature that various mechanisms, such as the formation of hydrogen peroxide by reduction of oxygen followed by the scission of $H_2O_2$ catalyzed by the presence of metal cations (*42*) or the simultaneous reaction of the two gaseous reactants $H_2$ and $O_2$ on the platinum surface of electrodes in an operating cell (*43*), can lead to the formation of the radicals $H^{\cdot}$ and $HO^{\cdot}$.

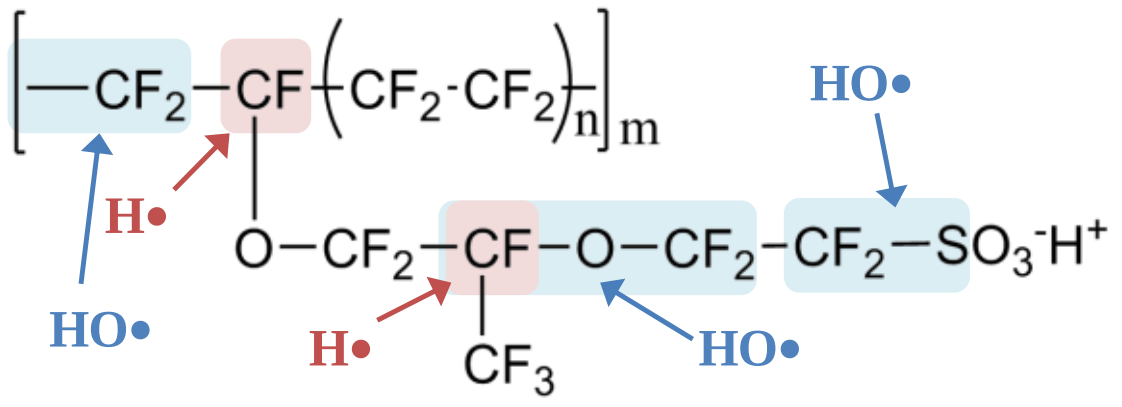

Figure 7. Schematic representation of possible radical attacks in Nafion during fuel cell operation, proposed by Ghassemzadeh et al. (*44*).

### 2.5 Membrane imaging and operando MRI

As NMR is non-invasive and selective, it is a method of choice for the operando study of water transfer in PEMFC cells. However, MRI is incompatible with ferromagnetic, paramagnetic or electrically conductive materials. Carbonated layers are key elements in the design of PEMFC cells and cannot be easily replaced due to their multiple roles in the system. The general advancement of MRI methods in the PEMFC research field has been limited in particular by the existence of carbon-based gas diffusion layers (GDLs). During the experiment, the radiofrequency (RF) waves generated by the excitation/reception radio-frequency coil must reach the measurement volume to excite the spins and collect the NMR signal in return. These signals are strongly attenuated by conductive materials, resulting in considerable losses and severe distortions and artifacts. Despite these disadvantages, MRI has been the target of much attention from the PEMFCs scientific community from 2004 until around 2012 (*45-52*). A major step in the field of operando MRI of PEMFCs has been achieved by the Balcom's group, who have used a prototype surface coil as an integrated radio-frequency resonator in the fuel cell prototype. In this configuration, the planar resonator serves simultaneously as a reactant supply plate for the cell and as a current collector (*52*). A probe of related design was introduced at the same time as this work for biomedical histology (*53*). The aim of this development was to increase NMR sensitivity and spatial resolution. Using this methodology, the authors were able to obtain one-dimensional water content profiles with a resolution of 6 $\mu m$ (*51*). The authors clearly capture the partial dehydration of the membrane on the anode side, even at room temperature. This very good spatial resolution was made possible here by the innovative design of the RF coil. The design avoids problems of RF field shielding by conductive layers (carbon electrodes and electrode support, gas diffusion layers), as the resonator generates a magnetic field parallel to the plane of the membrane-electrode assembly rather than normal to it, as is the case in "standard" configurations where the RF field is generated by a volume probe or surface coil. Although significant progress was made during the few years when MRI imaging of operating cells was being developed, the number of studies suddenly declined after 2012. It is still unrealistic to be able to carry out a time- and space-resolved MRI study of PEMFC operation under true operating conditions (particularly in terms of temperature) and on a prototype that enables realistic operation in terms of electrochemical performance. The incompatibility of materials rules out measurements on commercial PEMFC, and even the possibility of seeing the water in GDL, while the need for fast experiments with good spatial resolution means using small RF coils and a low field of view, which limits the size of the cell. Finally, and perhaps the most limiting factor, the thickness of the membranes currently in use is less than $20\ \mu m$. The resolution of MRI methods is therefore unsuitable for high-quality operando studies. That said, imaging methods can be used to study membrane properties ex-situ, outside the fuel cell system. This approach, detailed in the next paragraph, is the one that has been most widely followed in recent years for fundamental studies of diffusion in stressed membranes and water transfer at membrane interfaces.

## 3. NMR and MRI in a single proton exchange membrane

The strategy used for accessing NMR measurements on a single membrane involves the development of suitable radio-frequency (RF) NMR instrumentation coupled to dedicated devices, enabling the sample to be subjected to controlled mechanical and hydric stresses.

### 3.1 Water self-diffusion in a membrane under traction / compression

The emission of the RF field and the detection of the NMR signal are both performed by a copper wire loop of diameter $d = 10\ mm$. The dimensions are sufficiently small compared to the wavelength at $100\ MHz$. When the electronic circuit is tuned and matched by a symmetrical capacitive circuit, the coil has a good quality factor of around $Q \sim 110$. In this geometry, the plane of the membrane is close to the coil and therefore exposed to a strong RF magnetic field, ensuring a significant gain in signal-to-noise ratio. The plane of the membrane is located within the linearity zone of the RF field, i.e. at around $d/4 = 2.5\ mm$.

- For the traction experiments the dimensions of the membrane were chosen so that the coil covers part of the stretched sample as uniformly as possible (away from the ends). The mini-tensile machine was made out of polycarbonate and PTFE (Figure 8(a)). It consisted of two sliding jaws, which moved symmetrically with respect to the center of the sample. This means that, whatever the stretch ratio, the center of the sample always coincided with the center of the RF coil, the center of the field gradient coils, and the center of the magnet homogeneity zone.

- For the compression experiments, an NMR-compatible device was developed to quantify the evolution of the water content of a PFSA membrane exposed to a variable and balanced normal stress in a controlled relative humidity environment (Figure 8(b)). The body ($36 \times 44\ mm$) of the compression cell consists of a chamber, a movable piston, a lid and a chamber cover made of polyetheretherketone (PEEK). The $5\ mm$-diameter membrane sample is placed between two porous quartz discs (pore size $15$ to $40\ \mu m$) to allow water to drain away when pneumatic pressure is applied. Brass screws and nuts were used to tighten the assembly, with polymethyl methacrylate (PMMA) spacers. The NMR coil was positioned around the lower porous disk below the plane of the membrane. To maintain constant, homogeneous humidity conditions in the membrane environment, a flexible hose was placed close to the membrane between the two spacers to supply humid air at a fixed relative humidity ($15\% < RH < 98\%$).

The traction and compression systems were inserted in a Bruker Biospec 24/40 imager, equipped with a 3-axis field gradient sheath producing a maximum intensity of $20\ Gauss/cm$ ($0.2\ T/m$). The combination of two orthogonal field gradients in the $(xy)$ plane made it possible to modify the orientation of the magnetic field gradient relatively to the plane of the membrane, and thus control the direction of diffusion coefficient measurement. A stimulated spin-echo sequence (PGSTE) with unipolar magnetic field gradients was used to perform the experiments.

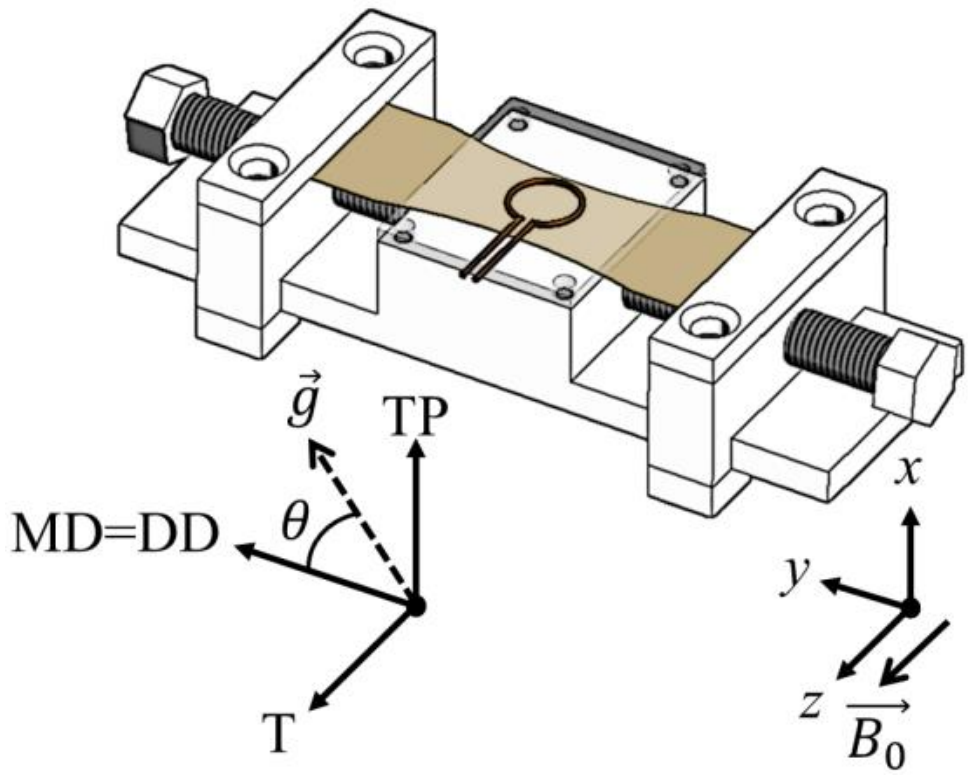

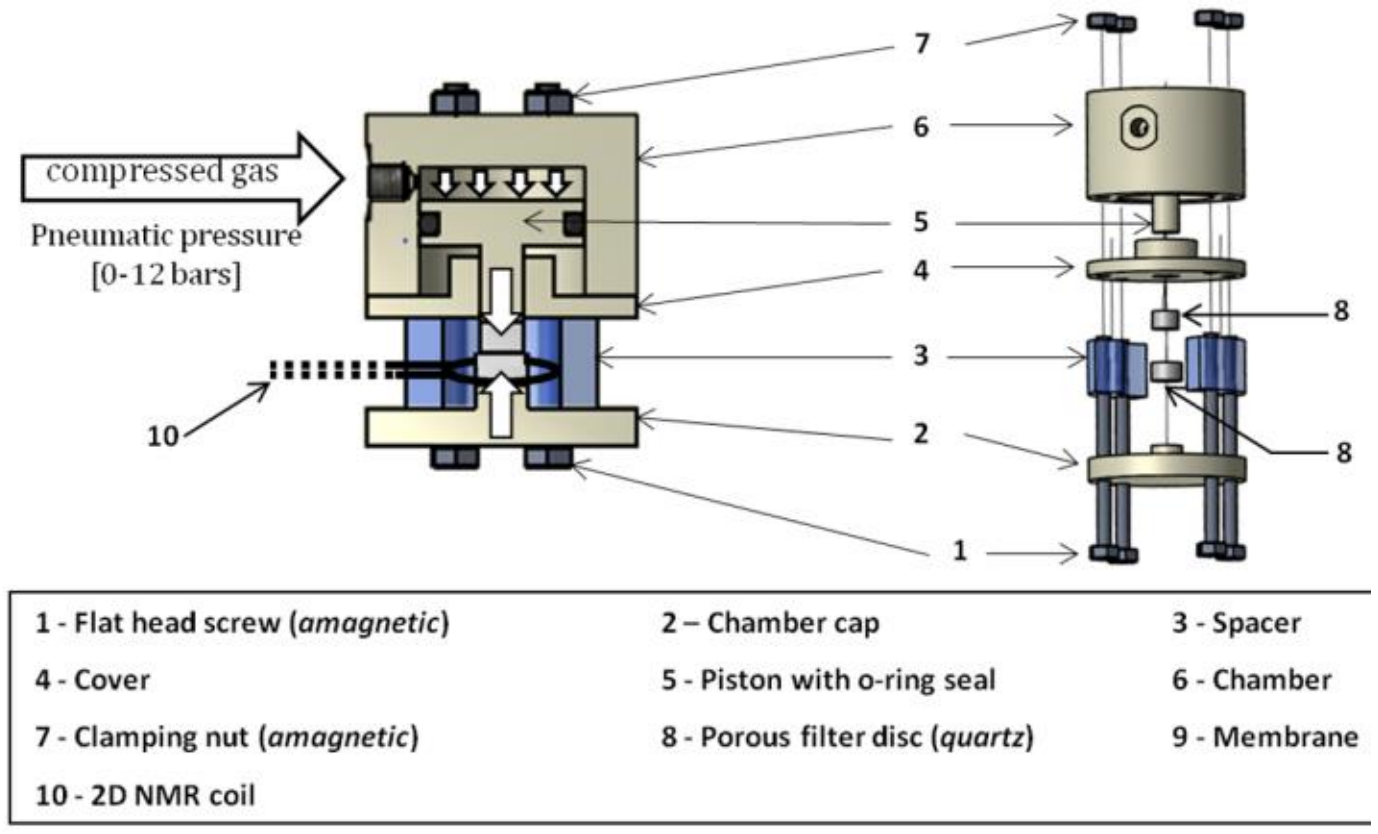

(a) (b)

Figure 8

(a) Mini-tensile machine and surface coil developed for measuring the water self-diffusion coefficient in membranes under tension. Figure from reference (*54*, *55*).

(b) Compression mini-machine and surface coil developed to characterize the effect of compressive stress on water diffusion properties. Figure from reference (*56*)

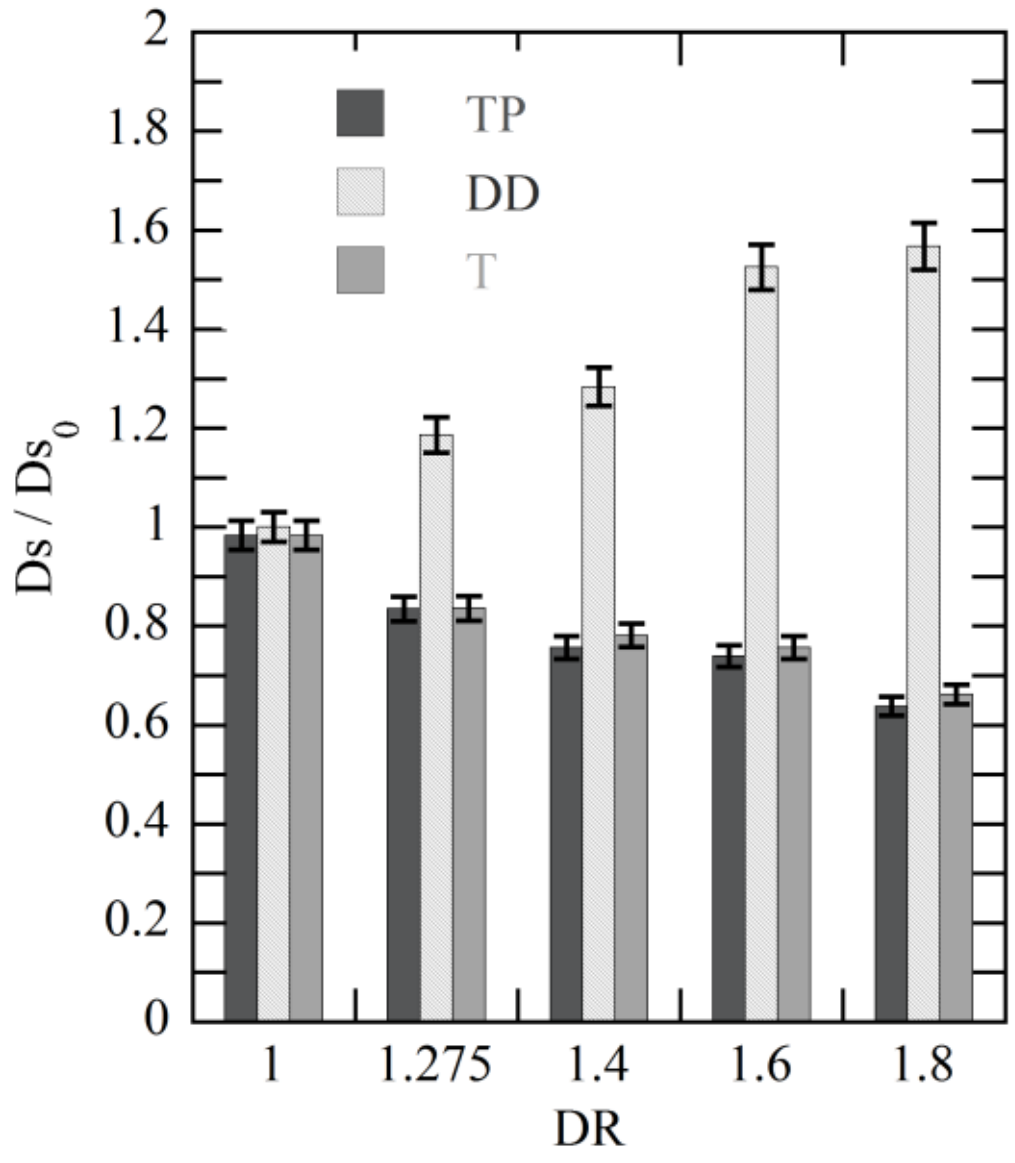

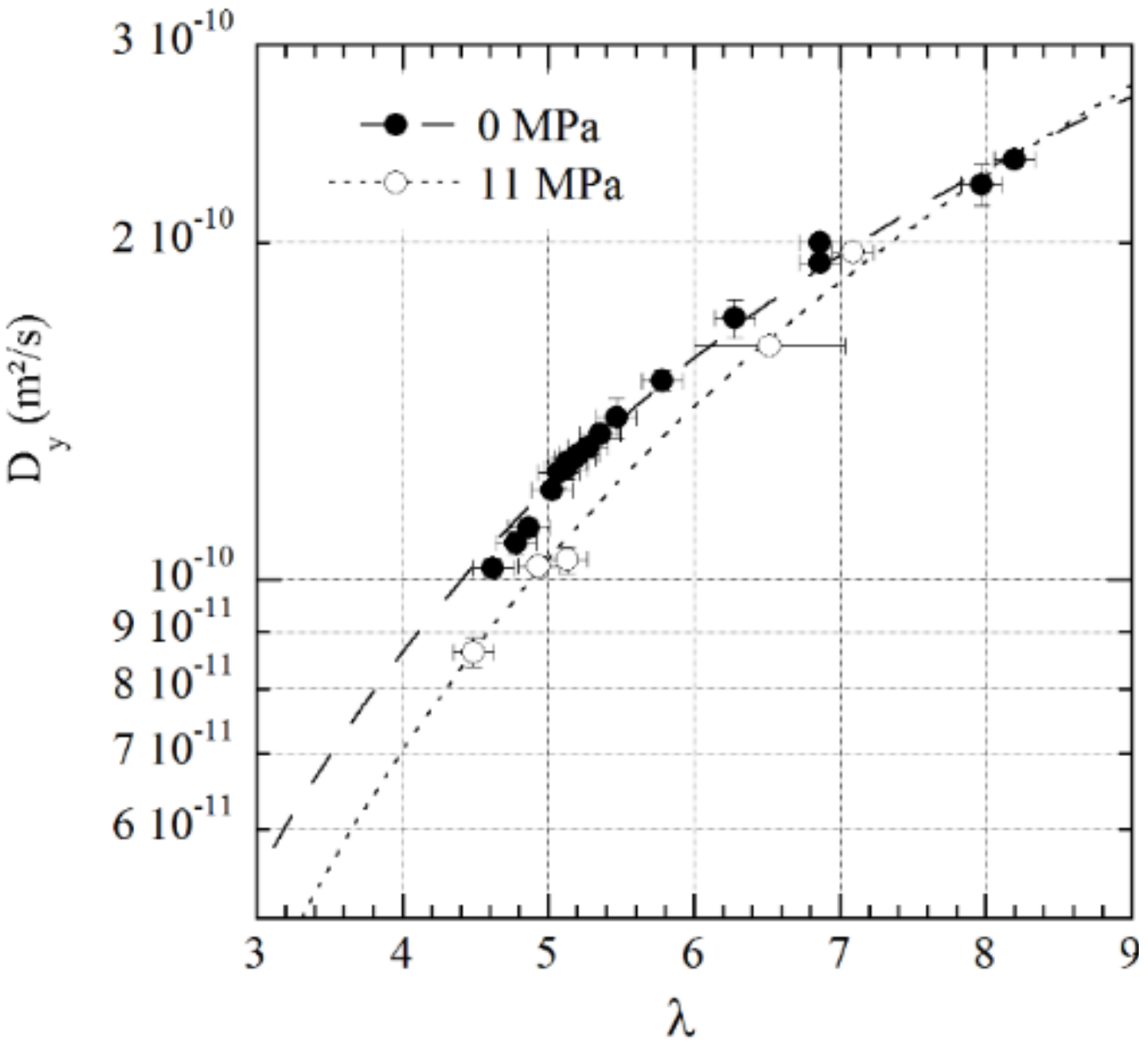

(a) (b)

Figure 9

(a) Water self-diffusion coefficient measured along the three main axes of the diffusion tensor: stretching direction ($DD$), through-plane direction ($TP$) and transverse direction ($T$). Values are normalized by the coefficient measured in the $DD$ direction in the unstretched sample. Figure from reference (*54*).

(b) Effect of compression on water diffusion, measured in the direction of stress application. Figure from reference (*56*).

Figure 9(a) shows the evolution of the diffusion coefficient in a single Nafion N1110 membrane ($250\ \mu m$ nominal thickness) equilibrated at $\lambda = 4 \pm 0.5$ for different stretching ratios. In the unstretched sample ($DR = 1$), the lamination process induces a weak alignment of the Nafion membrane structure, resulting in a $\sim 10\%$ anisotropy of the water diffusion coefficient. Before stretching, the aggregates of polymer chains, assembled in lateral size domains of around $800$ Å, are randomly oriented on a micrometric scale. A weak orientation order exists, however, in membranes that were laminated (N112, N115, N117, N1110), but not in those obtained by casting-evaporation, for which the order is orthogonal to the plane of the membrane. Measurements in the $DD$, $TP$ and $T$ directions for increasing draw ratios demonstrate that the diffusion tensor is cylindrically symmetrical around the stretching direction with excellent approximation. When stretched, the domains rotate in the tensile direction. The symmetry of the structure is uniaxial, as is that of the diffusion tensor. The diffusion anisotropy increases strongly to reach 2.5 for $DR =$ 1.7. Beyond this point, the domains deform, probably introducing an alignment of polymer chain aggregates along the tensile direction. At room temperature, systematic membrane rupture occurs at $DR \sim 2.2$ .

Measurements of the water self-diffusion coefficient in samples under compressive stress (Figure 9(b)) have enabled us to quantify the impact of compression on membrane structure and transport. The results show a clear drop in content and diffusion coefficient with increasing stress. The relative effect on diffusion is much greater than on water content, in agreement with the strong acceleration of diffusion at low water content (Figure 5 and Figure 9(b)). The effect of a change in microstructure is also experienced throughout the hydration range.

### 3.2 Water transfer at the membrane/humid air interface

Ex-situ membrane imaging can be effectively used to study phenomena taking place at the interfaces between the membrane and an external humid atmosphere. Among the phenomena of interest is the transfer of water from a gaseous phase to the membrane interior. The experiment which aims to subject a membrane to a flow of humid air on one side and a flow of dry air on the other corresponds to the gas-phase pervaporation process. It consists in three phases:

- water sorption in the membrane on the wet air side;
- diffusion of water through the structure;
- water desorption on the dry air side.

Sorption and desorption can be accompanied by interface resistance phenomena, controlled by the boundary layer in the gas flow, the gas ↔ liquid phase change and the surface properties (morphology, hydrophobicity) of the membrane (Figure *10*). While boundary layer effects can be minimized by imposing a high gas flow rate, the other two resistances cannot be avoided. However, their impact on the overall phenomenon is more or less important, depending on the relative values of the resistance due to diffusion inside the membrane $R_D$ and the interface resistances $R_{int}$.

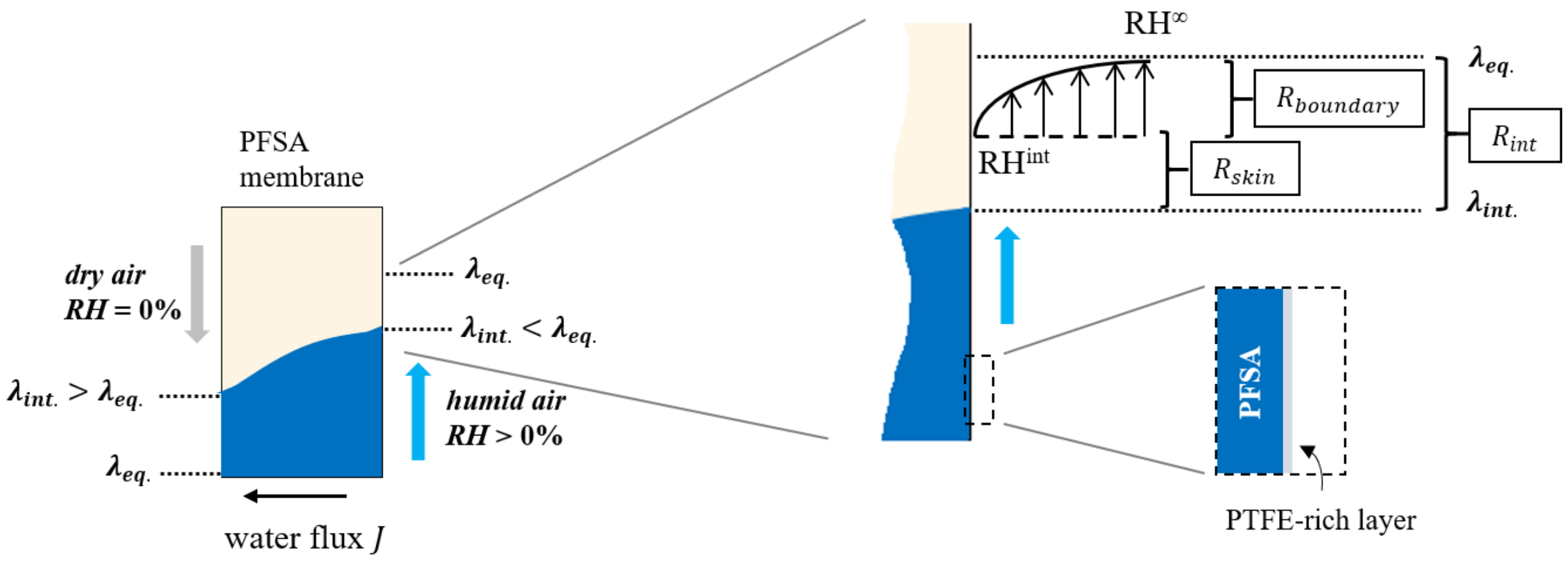

(a) (b)

Figure 10

(a) Illustration of the effect of an interfacial resistance on the steady-state water content at the interface. The non-linear water profile results from the concentration dependence of the diffusion coefficient.

(b) Zooming in on the interface reveals the presence of two resistances in series, one of which is due to the boundary layer effect and the other to the morphology of the thin PTFE-rich layer on the membrane surface.

Experimental results acquired when a Nafion N1110 membrane is subjected to a moisture gradient between its two faces are shown in Figure *11*. The water profile developed across the material under steady-state conditions is imaged by one-dimensional MRI. The method used (*57, 58*) is based on the work of Ouriadov et al (*59*), who demonstrated the use of a surface coil for MRI measurement of 1D concentration and relaxation time profiles in thin films. The sequence used is a Single Point Imaging (SPI) sequence including a spin echo (SE-SPI).

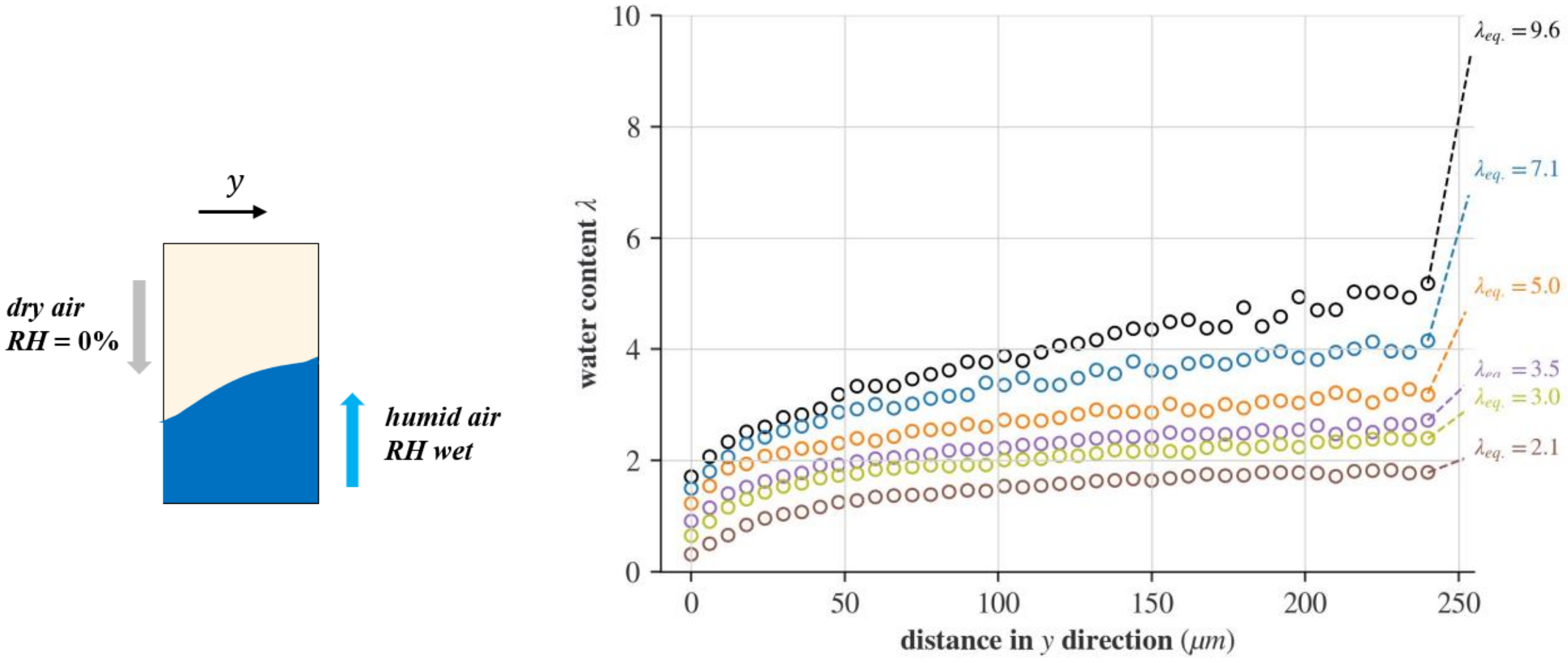

(a) (b)

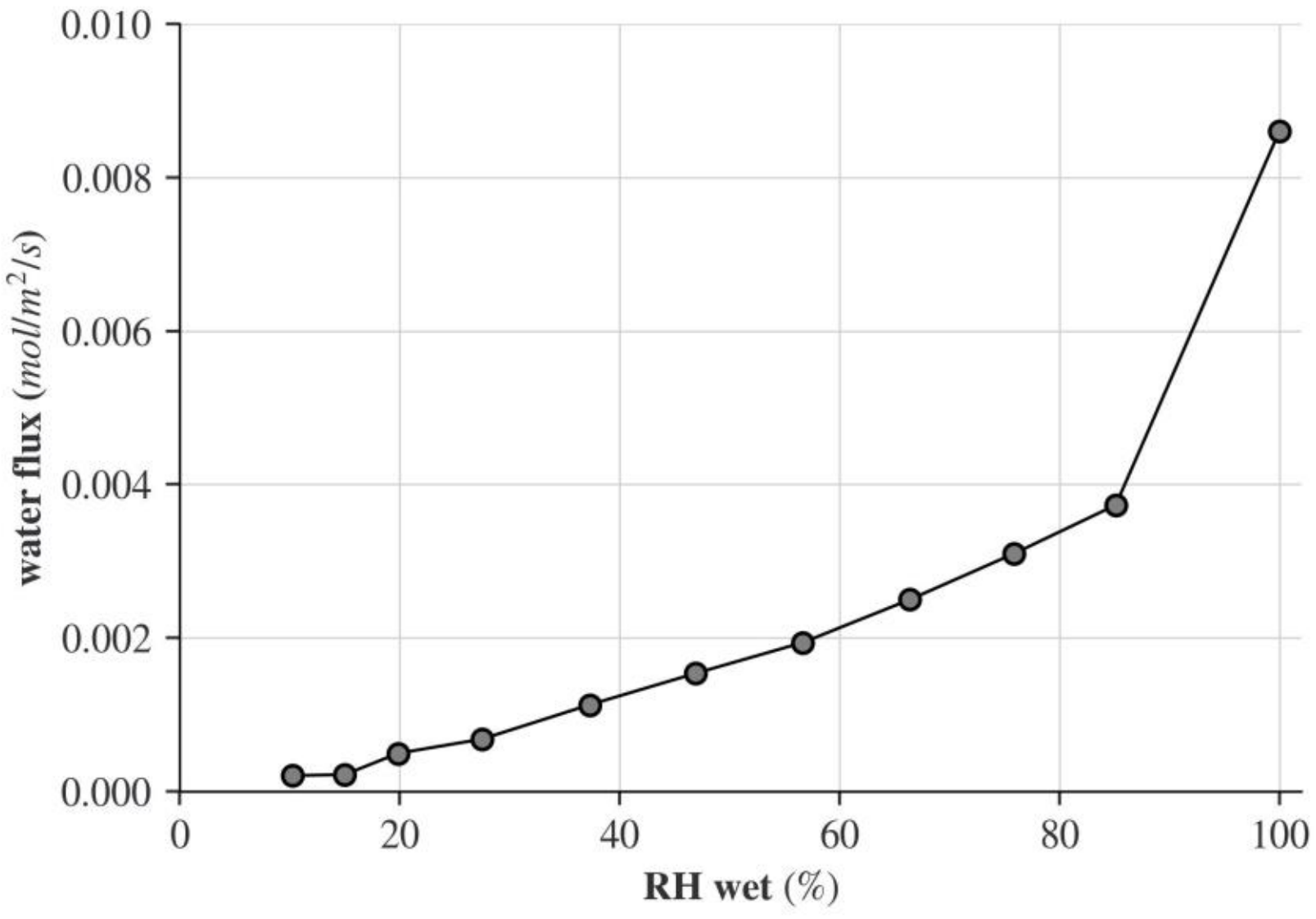

(c)

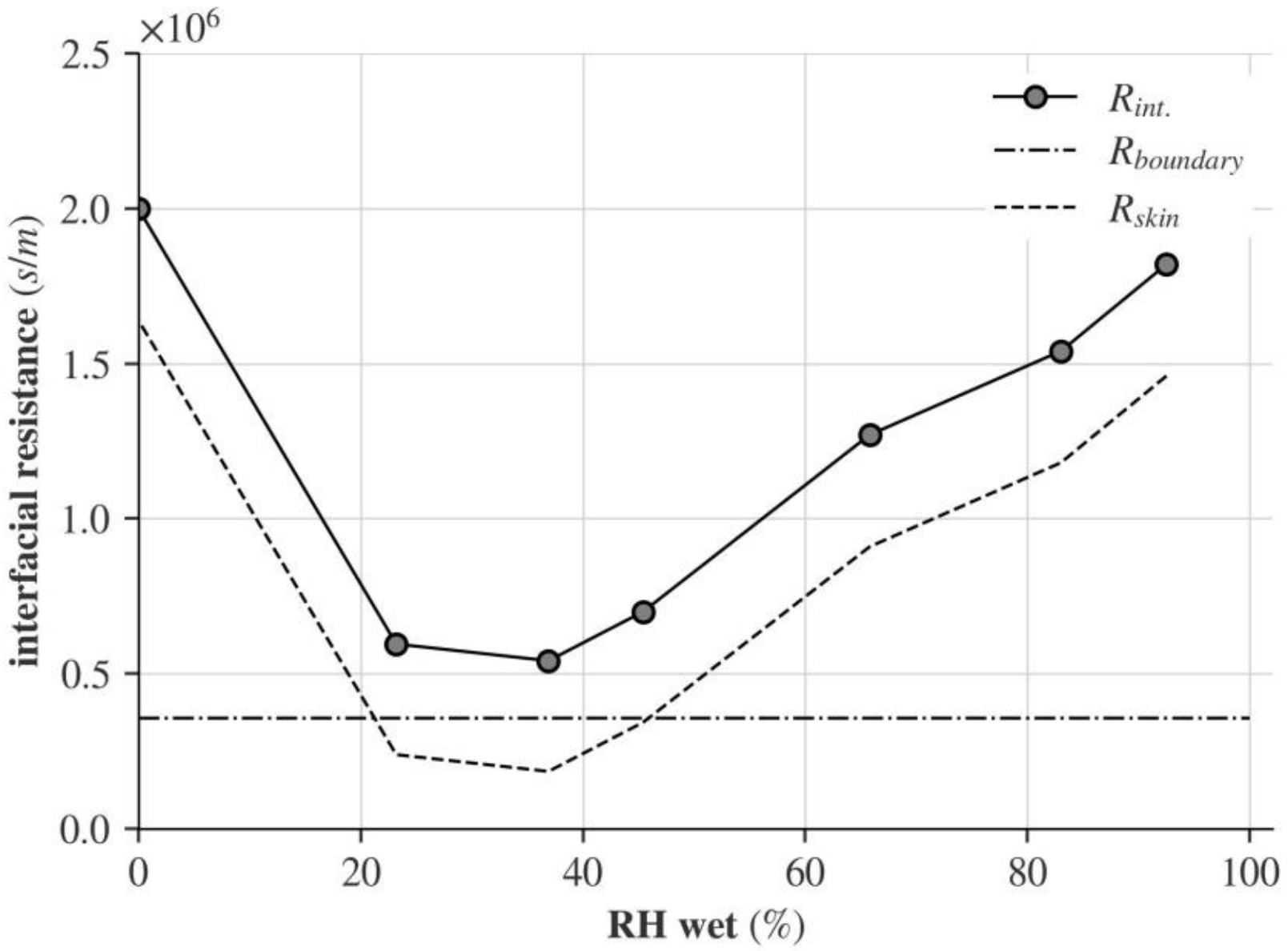

(d)

Figure 11

(a) Experimental situation: one side of the membrane is subjected to a flow of dry air, while the other is subjected to a flow of humid air. Relative humidity is varied in the range $[\sim 20\% - \sim 90\%]$. The water profiles measured by MRI (b) are recorded in the steady-state regime for each air humidity condition. (c) Water flux measured through the membrane from the wet to the dry side. (d) Evolution of the total interface resistance $R_{int}$, the constant resistance due to boundary layer effects ($R_{boundary}$) and the interfacial resistance due to membrane surface morphology $R_{skin}$ as a function of the air humidity imposed on the wet side.

An effective mass transfer coefficient $k$ can be defined as:

$$J = k[C_{int} - C_{\infty}] = k\left[\lambda_{int} - \lambda_{eq.}\right] \text{ with } k = \frac{h}{\gamma}\frac{P_{sat(T)}}{RT} \qquad \text{equation 2}$$

where $J$ is the water flux through the membrane, $C_{int}$ and $C_{\infty}$ are respectively the water concentration at the interface and the corresponding equilibrium water concentration at the given relative humidity, away from the surface ($RH^{\infty}$). $P_{sat}(T)$ is the saturation vapor pressure at temperature $T$, and $\gamma$ is the local slope of the sorption isotherm (*60*). The interface resistance is $R_{int} = 1/k$ and can therefore be deduced from the equilibrium concentration ($\lambda_{eq.}$ is given by the sorption curve), the water flux measurement ($J$ is deduced from RH probes measurements at the inlet and outlet of the cell containing the membrane) (Figure *11*(c)) and the water content at the interface ($\lambda_{int}$ is obtained from the MRI profiles, Figure *11*(b)). The decay measured at low humidity in the evolution curve of $R_{int}$ (Figure *11*(d)) can be rationalized based on morphological changes in the PFSA surface. Indeed, surface morphology adopts moisture-dependent configurations, with increased hydrophilicity and conduction channels for water. These channels are also more open and less tortuous as moisture increases (*1*). In the limiting case where the membrane surface is in equilibrium with liquid water, we anticipate a very sharp drop in interface resistance, down to a value close to zero (*61-63*). The water conduction channels are then predominantly oriented towards the surface, this configuration being the one that minimizes surface energy, and a water film forms and propagates towards the interior of the material. By subtracting the constant value of resistance due to limitations to mass transfer at the interface due to the boundary layer phenomenon $R_{boundary}$, we can determine an approximate value of resistance due to the morphological skin effect ($R_{skin}$). This effect, which appears here to dominate interface resistance over a wide humidity range, stems from the highly hydrophobic nature of the surface, comparable to that of Teflon. This condition is due to the pendant chains carrying the ionic sites, which are oriented at the interface towards the interior of the material over a thickness of a few nanometers (*64*). This results in a lower water content over a short distance. In conclusion, the limitations to steady-state water transfer in Nafion come mainly from interface effects when the condition is of the "humid air" type. When air humidity is low, resistance due to diffusion can be significant, as can that due to interface hydrophobicity. If the gas flow rate is sufficiently high, the boundary layer effect can be limited. In the transient regime, another limitation, arising from polymer relaxation under the effect of chain rearrangement associated with swelling, must be considered when modeling the overall phenomenon.

## 4. Conclusion

PEMFC fuel cells are electrochemical converters that play a significant role in the energy mix. Understanding the properties of proton exchange electrolyte membranes is critical for creating efficient and long-lasting materials. Advanced experimental methods are required for this. In this article, we reviewed a number of studies that investigated water behavior in Nafion reference membranes using NMR spectroscopy, field cycling NMR relaxometry, PFG-NMR, and MRI imaging. Over the year this membrane has evolved into a model system on which new approaches can be tested. Despite the large number of studies that have been published in the literature, research on the Nafion membrane is not yet complete. Studies are currently focused on understanding the membrane's degradation mechanisms and the strategies to be followed to limit them. Other studies are being carried out on alternative materials. One option is to introduce modifications into hydrocarbon host membranes, with the aim of giving them good ionic conduction properties and improved durability. The quest for the "ideal" membrane, which would have all the advantages of Nafion without its shortcomings, is therefore still ongoing.

## Funding

This research has received funding from the European Union's EIT Raw Materials project nº 19247 ALPE: "Advanced Low-Platinum hierarchical Electrocatalysts for low-T fuel cells".

## References

1. A. Kusoglu, A. Z. Weber, New Insights into Perfluorinated Sulfonic-Acid lonomers. *Chemical Reviews* **117**, 987-1104 (2017).
2. N. Cornet *et al.*, Sulfonated polyimide membranes: a new type of ion-conducting membrane for electrochemical applications. *Journal of New Materials for Electrochemical Systems* **3**, 33-42 (2000).
3. Y. Sone, P. Ekdunge, D. Simonsson, Proton conductivity of Nafion 117 as measured by a four-electrode AC impedance method. *Journal of the Electrochemical Society* **143**, 1254-1259 (1996).
4. K. D. Kreuer, S. J. Paddison, E. Spohr, M. Schuster, Transport in proton conductors for fuel-cell applications: Simulations, elementary reactions, and phenomenology. *Chemical Reviews* **104**, 4637-4678 (2004).
5. L. Rubatat, A. L. Rollet, G. Gebel, O. Diat, Evidence of elongated polymeric aggregates in Nafion. *Macromolecules* **35**, 4050-4055 (2002).
6. L. Rubatat, PhD thesis. Université Joseph Fourier, Grenoble I, (2003).
7. J. C. Perrin, S. Lyonnard, F. Volino, Quasielastic neutron scattering study of water dynamics in hydrated nafion membranes. *Journal of Physical Chemistry C* **111**, 3393-3404 (2007).
8. Q. A. Zhao, P. Majsztrik, J. Benziger, Diffusion and Interfacial Transport of Water in Nafion. *Journal of Physical Chemistry B* **115**, 2717-2727 (2011).
9. J. C. Perrin, S. Lyonnard, A. Guillermo, P. Levitz, Water dynamics in ionomer membranes by field-cycling NMR relaxometry. *Journal of Physical Chemistry B* **110**, 5439-5444 (2006).
10. J. C. Perrin, S. Lyonnard, A. Guillermo, P. Levitz, Water dynamics in lonomer membranes by field-cycling NMR relaxometry. *Fuel Cells* **6**, 5-9 (2006).
11. J. C. Perrin, S. Lyonnard, A. Guillermo, P. Levitz, Water dynamics in ionomer membranes by field-cycling NMR relaxometry. *Magnetic Resonance Imaging* **25**, 501-504 (2007).
12. E. L. Hahn, Spin Echoes. *Physical Review* **80**, 580-594 (1950).
13. W. S. Price, Pulsed-field gradient nuclear magnetic resonance as a tool for studying translational diffusion: Part 1. Basic theory. *Concepts Magn. Reson.* **9**, 299-335 (1997).
14. *Diffusion NMR of Confined Systems*. (Royal society of chemistry, 2017).
15. S. Suarez, S. Greenbaum, Nuclear Magnetic Resonance of Polymer Electrolyte Membrane Fuel Cells. *Chemical Record* **10**, 377-393 (2010).
16. L. M. Yan, Y. D. Hu, X. M. Zhang, B. H. Yue, Applications of NMR Techniques in the Development and Operation of Proton Exchange Membrane Fuel Cells. *Annual Reports on Nmr Spectroscopy, Vol 88* **88**, 149-213 (2016).
17. Z. Zhang, B. Balcom, *PEM fuel cell diagnostic tools*. Magnetic Resonance Imaging (CRC Press, 2017), pp. 229-254.
18. J.-C. Perrin, PhD thesis. Université Joseph Fourier, Grenoble I, (2006).
19. N. J. Bunce, S. J. Sondheimer, C. A. Fyfe, Proton NMR method for the quantitative determination of the water content of the polymeric perfluorosulfonic acid Nafion-H. *Macromolecules* **19**, 333-339 (1986).
20. C. Wakai, T. Shimoaka, T. Hasegawa, Analysis of the Hydration Process and Rotational Dynamics of Water in a Nafion Membrane Studied by H-1 NMR Spectroscopy. *Analytical Chemistry* **85**, 7581-7587 (2013).

21. L. Maldonado, J. C. Perrin, J. Dillet, O. Lottin, Characterization of polymer electrolyte Nafion membranes: Influence of temperature, heat treatment and drying protocol on sorption and transport properties. *Journal of Membrane Science* **389**, 43-56 (2012).
22. N. Bloembergen, E. M. Purcell, R. V. Pound, Relaxation effects in Nuclear Magnetic Resonance Absorption. *Phys. Rev.* **73**, 679-712 (1948).
23. B. MacMillan, A. R. Sharp, R. L. Armstrong, An nmr investigation of the dynamical characteristics of water absorbed in Nafion. *Polymer* **40**, 2471-2480 (1999).
24. B. MacMillan, A. R. Sharp, R. L. Armstrong, N.m.r. relaxation in Nafion - The low temperature regime. *Polymer* **40**, 2481-2485 (1999).
25. A. Guillermo, G. Gebel, H. Mendil-Jakani, E. Pinton, NMR and Pulsed Field Gradient NMR Approach of Water Sorption Properties in Nafion at Low Temperature. *Journal of Physical Chemistry B* **113**, 6710-6717 (2009).
26. C. Wakai, T. Shimoaka, T. Hasegawa, H-1 NMR Analysis of Water Freezing in Nanospace Involved in a Nafion Membrane. *Journal of Physical Chemistry B* **119**, 8048-8053 (2015).
27. R. Kimmich, *NMR Tomography Diffusometry Relaxometry*. (Springer: Berlin Heidelberg, 1997).
28. J. P. Korb, S. Xu, J. Jonas, Confinement effects on dipolar relaxation by translational dynamics of liquids in porous silica glasses. *Journal of Chemical Physics* **98**, 2411-2422 (1993).
29. M. Holz, S. R. Heil, A. Sacco, Temperature-dependent self-diffusion coefficients of water and six selected molecular liquids for calibration in accurate H-1 NMR PFG measurements. *Physical Chemistry Chemical Physics* **2**, 4740-4742 (2000).
30. T. A. Zawodzinski Jr *et al.*, Comparative study of water uptake by and transport through ionomeric fuel cell membranes. *J. Electrochem. Soc.* **140**, 1981-1985 (1993).
31. K. D. Kreuer, On the development of proton conducting materials for technological applications. *Solid State Ionics* **97**, 1-15 (1997).
32. A. F. Privalov, V. V. Sinitsyn, M. Vogel, Transport Mechanism in Nafion Revealed by Detailed Comparison of 1H and 17O Nuclear Magnetic Resonance Diffusion Coefficients. *Journal of Physical Chemistry Letters* **14**, 9335-9340 (2023).
33. X. Gong *et al.*, Self-diffusion of water, ethanol and decafluropentane in perfluorosulfonate ionomer by pulse field gradient NMR. *Polymer* **42**, 6485-6492 (2001).
34. M. Zaton, J. Roziere, D. J. Jones, Current understanding of chemical degradation mechanisms of perfluorosulfonic acid membranes and their mitigation strategies: a review. *Sustainable Energy & Fuels* **1**, 409-438 (2017).
35. L. Dubau *et al.*, A review of PEM fuel cell durability: materials degradation, local heterogeneities of aging and possible mitigation strategies. *Wiley Interdisciplinary Reviews-Energy and Environment* **3**, 540-560 (2014).
36. M. Robert, A. El Kaddouri, J. C. Perrin, S. Leclerc, O. Lottin, Towards a NMR-Based Method for Characterizing the Degradation of Nafion XL Membranes for PEMFC. *Journal of the Electrochemical Society* **165**, F3209-F3216 (2018).
37. L. Dubau *et al.*, Carbon corrosion induced by membrane failure: The weak link of PEMFC long-term performance. *International Journal of Hydrogen Energy* **39**, 21902-21914 (2014).
38. G. De Moor *et al.*, Perfluorosulfonic acid membrane degradation in the hydrogen inlet region: A macroscopic approach. *International Journal of Hydrogen Energy* **41**, 483-496 (2016).

39. J. H. Han, K. W. Lee, C. E. Lee, H-1 nuclear magnetic resonance study of low-temperature water dynamics in a water-soaked perfluorosulfonic acid ionomer Nafion film. *Solid State Communications* **250**, 28-30 (2017).
40. A. Panchenko *et al.*, In-situ spin trap electron paramagnetic resonance study of fuel cell processes. *Physical Chemistry Chemical Physics* **6**, 2891-2894 (2004).
41. L. Lin, M. Danilczuk, S. Schlick, Electron spin resonance study of chemical reactions and crossover processes in a fuel cell: Effect of membrane thickness. *Journal of Power Sources* **233**, 98-103 (2013).
42. A. Pozio, R. F. Silva, M. De Francesco, L. Giorgi, Nafion degradation in PEFCs from end plate iron contamination. *Electrochimica Acta* **48**, 1543-1549 (2003).
43. V. O. Mittal, H. R. Kunz, J. M. Fenton, Membrane degradation mechanisms in PEMFCs. *Journal of the Electrochemical Society* **154**, B652-B656 (2007).
44. L. Ghassemzadeh, K. D. Kreuer, J. Maier, K. Muller, Chemical Degradation of Nation Membranes under Mimic Fuel Cell Conditions as Investigated by Solid-State NMR Spectroscopy. *Journal of Physical Chemistry C* **114**, 14635-14645 (2010).
45. J. Bedet *et al.*, Magnetic resonance imaging of water distribution and production in a 6 cm(2) PEMFC under operation. *International Journal of Hydrogen Energy* **33**, 3146-3149 (2008).
46. S. Tsushima, K. Teranishi, S. Hirai, Magnetic resonance imaging of the water distribution within a polymer electrolyte membrane in fuel cells. *Electrochemical and Solid State Letters* **7**, A269-A272 (2004).
47. K. R. Minard, V. V. Viswanathan, P. D. Majors, L. Q. Wang, P. C. Rieke, Magnetic resonance imaging (MRI) of PEM dehydration and gas manifold flooding during continuous fuel cell operation. *Journal of Power Sources* **161**, 856-863 (2006).
48. M. T. Wang, K. W. Feindel, S. H. Bergens, R. E. Wasylishen, In situ quantification of the in-plane water content in the Nafion (R) membrane of an operating polymer-electrolyte membrane fuel cell using H-1 micro-magnetic resonance imaging experiments. *Journal of Power Sources* **195**, 7316-7322 (2010).
49. K. W. Feindel, S. H. Bergens, R. E. Wasylishen, Use of hydrogen-deuterium exchange for contrast in (1)H NMR microscopy investigations of an operating PEM fuel cell. *Journal of Power Sources* **173**, 86-95 (2007).
50. Z. W. Dunbar, R. I. Masel, in *Proton Exchange Membrane Fuel Cells 8, Pts 1 and 2,* T. Fuller *et al.*, Eds. (2008), vol. 16, pp. 1001-1008.
51. Z. Zhang *et al.*, Magnetic resonance imaging of water content across the Nafion membrane in an operational PEM fuel cell. *Journal of Magnetic Resonance* **193**, 259-266 (2008).
52. Z. H. Zhang *et al.*, Zero-mode TEM parallel-plate resonator for high-resolution thin film magnetic resonance imaging. *Canadian Journal of Chemistry-Revue Canadienne De Chimie* **89**, 745-753 (2011).
53. M. D. Meadowcroft *et al.*, Direct magnetic resonance imaging of histological tissue samples at 3.0T. *Magnetic Resonance in Medicine* **57**, 835-841 (2007).
54. M. Klein *et al.*, Anisotropy of Water Self-Diffusion in a Nafion Membrane under Traction. *Macromolecules* **46**, 9259-9269 (2013).
55. J. C. Perrin *et al.*, in *Polymer Electrolyte Fuel Cells 13,* H. A. Gasteiger *et al.*, Eds. (2013), vol. 58, pp. 781-788.

56. A. El Kaddouri *et al.*, Impact of a Compressive Stress on Water Sorption and Diffusion in Ionomer Membranes for Fuel Cells. A H-1 NMR Study in Vapor Equilibrated Nafion. *Macromolecules* **49**, 7296-7307 (2016).
57. M. Klein *et al.*, in *Polymer Electrolyte Fuel Cells 13*, H. A. Gasteiger *et al.*, Eds. (2013), vol. 58, pp. 283-289.
58. C. Mrad *et al.*, NMR characterization of proton exchange membranes in controlled hygrometry conditions. *Journal of Membrane Science* **688**, (2023).
59. A. V. Ouriadov, R. P. MacGregor, B. J. Balcom, Thin film MRI - high resolution depth imaging with a local surface coil and spin echo SPI. *Journal of Magnetic Resonance* **169**, 174-186 (2004).
60. S. Didierjean *et al.*, Theoretical evidence of the difference in kinetics of water sorption and desorption in Nafion (R) membrane and experimental validation. *Journal of Power Sources* **300**, 50-56 (2015).
61. M. B. Satterfield, J. B. Benziger, Non-fickian water vapor sorption dynamics by nafion membranes. *Journal of Physical Chemistry B* **112**, 3693-3704 (2008).
62. C. W. Monroe, T. Romero, W. Merida, M. Eikerling, A vaporization-exchange model for water sorption and flux in Nafion. *Journal of Membrane Science* **324**, 1-6 (2008).
63. P. Majsztrik, A. Bocarsly, J. Benziger, Water Permeation through Nafion Membranes: The Role of Water Activity. *Journal of Physical Chemistry B* **112**, 16280-16289 (2008).
64. G. S. Hwang, D. Y. Parkinson, A. Kusoglu, A. A. MacDowell, A. Z. Weber, Understanding Water Uptake and Transport in Nafion Using X-ray Microtomography. *Acs Macro Letters* **2**, 288-291 (2013).